%% file: main.tex
\documentclass[preprint,3p,times]{elsarticle}

\usepackage{amssymb}
\usepackage{amsmath}

\input{includes-and-macros}

\journal{The journal of Artificial Intelligence (AIJ)}

\begin{document}

\begin{frontmatter}



\title{Coalition Tactics: Bribery and Control in Parliamentary Elections}


\author{Hodaya Barr, Eden Hartman, Yonatan Aumann, Sarit Kraus} 
\affiliation{organization={Bar-Ilan University},
            country={Israel}}

\begin{abstract}
    Strategic manipulation of elections is typically studied in the context of promoting individual candidates.
    In parliamentary elections, however, the focus shifts: voters may care more about the overall governing coalition than the individual parties' seat counts.
    This paper studies this new problem: manipulating parliamentary elections with the goal of promoting the collective seat count of a coalition of parties. 
    We focus on proportional representation elections, and consider two variants of the problem; one in which the sole goal is to maximize the total number of seats held by the desired coalition, and the other with a dual objective of both promoting the coalition and promoting the relative power of some favorite party within the coalition. 
    
    We examine two types of strategic manipulations:
    \emph{bribery}, which allows modifying voters' preferences, and \emph{control}, which allows
    changing the sets of voters and parties. 
    We consider multiple bribery types, presenting polynomial-time algorithms for some, while proving NP-hardness for others.
    For control, we provide polynomial-time algorithms for control by adding and deleting voters. In contrast, control by adding and deleting parties, we show, is either impossible (i.e., the problem is immune to control) or computationally hard, in particular, W[1]-hard when parameterized by the number of parties that can be added or deleted.
\end{abstract}






\end{frontmatter}



\section{Introduction}
Strategic election manipulation by way of bribery and control has been widely studied (see, e.g. \cite{betzler2009parameterized,elkind2009swap,elkind2020algorithms,faliszewski2016control,faliszewski2006complexity,faliszewski2009hard,keller2018approximating, liu2009parameterized,tao2022hard,Yang2019Complexity,zhou2020parameterized}). 
This large body of literature focuses on single-winner elections, e.g.~ presidential elections and elections for individual seats in Congress. In practice, however, many countries around the globe operate under parliamentary systems employing \emph{proportional representation} (of one form or another). In such systems, multiple parties are elected and operate together in the parliament.  Moreover, in many, if not most cases, no single party has sufficient power to govern on its own, and the parliament's operation rests on \emph{coalitions} of parties. Any attempt to manipulate multi-party parliamentary elections must take into account such party and coalition dynamics. Indeed, promoting a specific party may be of little significance if the party ends up outside the ruling coalition. 


As such, existing models of bribery and control are inadequate for parliamentary elections, as they ignore the rich dynamics of parliamentary politics.  It is rather interesting that while parliamentary systems are prevalent around the globe, the extensive literature on bribery and control has mostly ignored these systems. This paper takes the first steps toward filling this gap.


\paragraph{Our Contributions}
In this paper, we introduce and study strategic manipulations for promoting \emph{coalitions of parties} in parliamentary elections, focusing on \emph{bribery} --- which allows modifying voters’ preferences --- and \emph{control} --- which allows changing the sets of voters and parties. 
Our first contribution is a formal definition of the model and problems, introducing two variants of the problem. 
In the first variant, the manipulator's sole goal is to promote an entire coalition of parties.  In the second variant, we acknowledge that the manipulator may also be interested in promoting a specific party, and thus define a dual-objective problem --- both promoting an entire coalition and promoting a specific preferred party therein. Throughout, we focus on proportional representation elections, wherein each voter casts a single vote for one party.

Having defined the problem(s), we study the complexity of bribery and control in various settings, providing both algorithmic solutions and hardness results. 

For bribery, we show that the complexity depends on the cost structure of the bribery.  For the 1 and \$ cost structures~\cite{faliszewski2006complexity} the problem is polynomial-time solvable.  For the Swap and Shift cost structures~\cite{elkind2009swap} the problem remains polynomial-time solvable if there is no electoral threshold, but becomes NP-hard when such a threshold is introduced. Table \ref{tab:sum_complexity_bribery} summarizes the results for bribery.  

For control, we consider both voter control (adding or deleting voters) and party control (adding or deleting parties). 
Voter control, we show,  is polynomial-time solvable. 
For party control, we show that a few variants are immune to manipulation (i.e., no such manipulation can advance the manipulator's goals) and all other variants are provably hard.  That being the case, we also consider the parametrized complexity of the problem, where the parameter is the number of parties to add or delete.  We show that even under parametrized analysis, the hard variants remain hard, either W[1]-hard or W[2]-hard. Table \ref{tab:control-results-general} summarizes the results for party control.

\paragraph{Organization}

\Cref{sec:model} presents the model and definitions. 
\Cref{sec:bribery} focuses on bribery. 
\Cref{sec:voter-control} considers voter control. 
\Cref{sec:party-control} studies party control.
Section \ref{sec:conclusion} concludes with some future work directions.
To enhance readability, for some results we include only intuitive proof sketches in the main body of the paper and provide the full versions in the appendix.

\subsection{Related Work}

\paragraph{Single-Winner Elections} In the single-winner setting, the complexity of strategic manipulations such as bribery and control has been widely studied. The bribery problem was introduced by \citet{faliszewski2006complexity}, and has since received significant attention (e.g., \cite{bartholdi1992hard, brelsford2008approximability, faliszewski2008Nonuniform, faliszewski2015complexity, maushagen2022complexity, parkes2012complexity}). The control problem, first studied by \citet{bartholdi1992hard}, has similarly been the subject of extensive research (e.g., \cite{menton2013control, procaccia2007multi, put2016complexity, sina2015adapting, maushagen2020last}). 
See \citet{faliszewski2016control} for a comprehensive survey of both.

In single-winner settings, the goal is to ensure that a particular candidate wins, while here the objective is to promote a \emph{group} of candidates (parties). 
This difference increases the difficulty of the problem
as there may be more ways to increase the coalition’s total support. Determining whether a successful manipulation exists may require examining a large space of possibilities.

\paragraph{Multi-Winner Elections} 
Strategic manipulations and control in multi-winner elections were first proposed by \citet{meir2008complexity}, and have been extensively studied since (e.g.,  \cite{aziz2015computational,bredereck2016complexity,bredereck2021coalitional,obraztsova2013manipulation, Yang2019Complexity,Yang2025Destructive}.  In the multi-winner model, the outcome of the election is a \emph{set} of winning candidates, rather than a single winner.  The objective of the manipulation is to promote some property of this winning set, e.g.~ that it coincide with some desired set~\cite{aziz2015computational}, to maximize the sum of utilities of the winning candidates exceed some threshold \cite{meir2008complexity}, to maximize the intersection of the winning set with some desired set \cite{Yang2019Complexity} (and other related objectives therein), that some set of undesired candidates not win \cite{Yang2025Destructive}.   

Our setting is fundamentally different.  In our setting, there are no \emph{winners}. Rather, all candidates (that pass the electoral threshold) are represented in the parliament, but in proportion to their vote count, and the goal of the manipulation relates to these proportions - namely, that the coalition and preferred party receive \emph{sufficient representation}.
As such, the models are not directly comparable.


\paragraph{Weighted Voting Games}
Parliamentary elections can be viewed as a special case of weighted voting games \cite{taylor2021simple}, for which there is a growing body of research on strategic manipulating power indexes (such as Shapley-Shubik or Banzhaf) --- e.g., \cite{Aziz2011False, rey2014false, rey2018structural, zick2011shapley, zuckerman2012manipulating}. 
Although related, our setting is fundamentally different.
In traditional weighted voting games, the focus is on coalition formation \emph{after} the fact. In our terms, each party has a \emph{fixed} number of votes, and strategic manipulations (such as false-name splits) aim to increase the power of a given party during coalition formation (for example, to secure a greater number of key ministries).
In our setting, however, we assume that coalitions are fixed in advance through known political agreements. The objective is to increase the total vote count of a preferred coalition (for example, to attain a parliamentary majority and form the ruling government).

\paragraph{Parliamentary Elections and Strategic Voting}

To the best of our knowledge, \citet{slinko2010proportional} were the first to study strategic behavior in parliamentary elections.
However, they focus on \emph{strategic voting}.
In contrast to our work, research in this area analyzes voters' individual incentives and the equilibrium behaviors that emerge when voters act strategically.
There is a rich literature on strategic voting in parliamentary elections --- e.g., \cite{bowler2010strategic,cox2018portfolio,gschwend2016drives,MCCUEN2010316,meffert2010strategic}.

\paragraph{Promoting a Single Party}
Closest to ours is the work by \citet{put2016complexity}, who also study the computational complexity of various strategic manipulations in parliamentary elections. 
However, their analysis focuses on promoting a \emph{single} party, whereas our goal is to promote an entire \emph{coalition}. In that sense, their problem can be seen as a special case of ours, where the coalition consists of only one party.
This difference once again introduces a combinatorial challenge: when promoting a coalition, there might be many more possible ways to affect the total support. Thus, additional techniques are needed.
We consider all the manipulations studied in their paper, as well as additional ones.
We recommend their related work section for further discussion.






\section{Model}\label{sec:model}






A parliamentary election is a triplet $E:=(P, V,\bm{\succ})$, where $P:=  \{p_1, \ldots, p_m\}$ is the set of \emph{parties}, 
$V:= \{v_1,\ldots,v_n\}$ is the set of \emph{voters}, and $\bm{\succ}:=(\succ_{v_1}, \ldots, \succ_{v_n})$ are the voters' strict \emph{preference orders} over $P$.
We denote by $\pos{E}{v_i}{p}$ the position of party $p$ according to $\oi_{v_i}$, and by $\topT{E}{v_i}$ the most preferred party according to $\oi_{v_i}$.  We shorthand $\oi_i$ for $\oi_{v_i}$ whenever the meaning is clear from the context.


Each voter contributes a single vote allotted to voter's most-preferred party participating in the election $E$.
The total number of votes allotted to party $p$ is denoted by $\Npoint{E}{p}$.
The fraction of votes received by party $p$ out of the total number of votes is denoted by $\seats{}{E}(p) := \Npoint{E}{p}/n$.

\paragraph{Electoral-threshold}
To avoid a parliament with small parties, many electoral systems impose a threshold $\thrs$, called the \emph{electoral-threshold}, such that only parties with $\seats{p}{}\geq \thrs$, get to be represented in parliament. These parties are called \emph{active parties}, and the set thereof is $P_A$. Votes to active parties are \emph{active votes}.

Lastly, the \emph{fraction of active votes} obtained by party $p$ is:
\begin{align*}\displaystyle
\seats{\thrs}{E}(p) := \begin{dcases}
			\frac{\Npoint{E}{p}}{\sum_{p' \in P_A} \Npoint{E}{p'}} & \text{if $p \in P_A$}\\
            0 & \text{otherwise}
		 \end{dcases}
\end{align*}
The function $\seats{\thrs}{E}(\cdot)$ is naturally extended to sets, $\seats{\thrs}{E}(A)=\sum_{p\in A}\seats{\thrs}{E}(p)$.



\subsection{Manipulation}
An external manipulator may modify the original election $E$ into another election $\hat{E} := (\hat{P}, \hat{V}, \hat{\bm{\succ}})$, where the possible modifications will be detailed in the respective chapters.  
This manipulator is referred to as the \emph{briber} in the context of bribery, and as the \emph{chair} in the context of control.

We assume there is a set $C$ of parties (actually running in $E$ or not), which we call the \emph{coalition}.   
The manipulator aims to increase the fraction of active votes of the coalition, and possibly also those of some  \emph{favored party} therein.  Specifically, the manipulator aims to construct a modified election instance $\hat{E}=(\hat{P},\hat{V},\hat{\bm{\succ}})$ satisfying one or both of the following objectives:

\begin{definition}[Joint Objective]  
Given \emph{coalition target-fraction} $\varphi \in (0,1]$, the objective is that the fraction of active votes received by the coalition parties is at least $\varphi$:
    \begin{align}\label{eq:target-coalition-obj}
        \tag{OBJ-J}
        \seats{\thrs}{\hat{E}} (C) \geq \varphi
    \end{align}
\end{definition}

\begin{definition}[Favored-Party Objective]  
For a \emph{favored-party}, $p_1 \in C$, and a \emph{target-favored-party-ratio} $\rho \in [0,1]$, the objective is that the fraction of active votes received by $p_1$, relative to the votes of the coalition, is at least $\rho$:
     \begin{align}\label{eq:target-favored-party-obj}
         \tag{OBJ-F}
         \seats{\thrs}{\hat{E}}(\prefferedParty) \geq \rho \cdot  \seats{\thrs}{\hat{E}}(C) 
     \end{align}
\end{definition}
We separately consider two settings: where the manipulator aims to satisfy both objectives, a setting we denote by \textbf{J+F}, and the setting where the manipulator aims to satisfy only the Joint objective, a case we denote by \textbf{J}.


Clearly, \textbf{J} is a special case of \textbf{J+F} (by setting $\rho = 0$). Thus, any hardness result for the former implies hardness for the latter, and any polynomial-time algorithm for the latter applies to the former as well.

\paragraph{The Proportion Objective} In our model, the manipulator's goals are expressed purely in terms of the \emph{fraction} of active votes awarded to the coalition and the preferred party. In reality, parliamentary systems and their elections are often considerably more complex. 

In Germany, for example, 299 Bundestag seats are filled via single-winner regional contests, while another 299 are determined through nationwide proportional representation (using the Sainte-Laguë/Schepers method). Subsequently, \emph{additional} seats --- known as \emph{Leveling Seats} --- are allocated to rectify any disproportionality between national vote share and seat count. In South Korea, 253 members are elected via single-winner districts, and 47 through national proportional representation; these are further divided into 17 ``Traditional'' PR seats and 30 ``Compensatory'' seats. Each country, thus, has its own specific electoral mechanics. 

In this paper, we abstract away from these institutional details (including specific apportionment methods), and use vote fractions as a proxy for parliamentary power. While admittedly a simplification, we argue that this serves as a sufficiently accurate proxy for the purposes of our analysis. We further note that under the Jefferson--D’Hondt apportionment method, seat shares have been shown, both empirically and theoretically, to be an affine function of vote shares~\cite{boratyn2024seatallocationseatbias,FlisEtAl2020}.

\paragraph{Naming Nomenclature} Problems are denoted using the format: 
\problemdef{\emph{type}}{\emph{sub-type}}{\emph{objective}}{\emph{threshold}},
where:
\begin{itemize}
    \item \emph{Type:} \textbf{B} for Bribery or \textbf{C} for Control.
    \item \emph{Sub-type:} defined in the respective sections.
    \item \emph{Objective:} J or J+F.
    \item \emph{Threshold:} $\thrs=0$ or $\thrs\geq 0$.
\end{itemize}
For example, \problem{B}{\$}{J+F}{\geq} denotes the \textbf{B}ribery problem, the sub-type is \$, the objectives are both \eqref{eq:target-coalition-obj} and \eqref{eq:target-favored-party-obj}, and the electoral threshold is $\thrs\geq 0$.

\paragraph{Notations}
We denote the coalition without the favorite party by $C_{-1}=C\setminus\{p_1\}$ and the opposition by $\bar{C}=P\setminus C$.  
Further, we denote the minimum number of votes necessary for passing the electoral-threshold by $T:= \lceil \tau \cdot |V| \rceil$. 

\paragraph{Example} 
As an example, consider an election with 5 parties $P\!=\!\{ p_1,c_1,c_2,o_1,o_2\}$, with the coalition $C\!=\!\{ p_1,c_1,c_2\}$.
There are 75 voters that can be grouped into five \emph{types}, distinguished by their preference orders, as follows: 

\begin{center}
\begin{tabular}{|c|c|c|c|c|c|}
\hline
 \diagbox[height=0.7cm, innerrightsep=5mm]{}{Type}& 1 & 2 & 3 & 4 & 5 \\
\hline
Top choice & $p_1$ & $c_1$ & $o_1$ & $o_2$ & $o_1$ \\
\hline
Second choice & $o_1$ & $o_1$ & $c_1$ & $c_2$ & $p_1$ \\
\hhline{|=|=|=|=|=|=|}
Number of voters & 20 & 10 & 20 & 20 & 5 \\
\hline
\end{tabular}
\end{center}
Suppose that the election threshold is $\thrs=15\%$, and both the coalition target-fraction and the target-favored-party-ratio are $\varphi=\rho=50\%$. 
Then, without manipulation, from the coalition only $p_1$ has active votes and $\seats{\thrs}{E}(C)=\seats{\thrs}{E}(p_1)\approx 31\%$, which is less than the required target of $50\%$ --- see the first row of Table~\ref{tab:example}. 
%
%

Suppose the manipulator can remove one (and only one) party.  
Table \ref{tab:example} summarizes the effect of removing the different parties. 
Note that $p_1$ gains the most from removing $o_1$, as this increases its share to $33\%$.  This removal also increases the coalition's share to $73\%$.  However, it reduces $p_1$'s \emph{relative} share within the coalition to $45\%$, which is below the required ratio of $50\%$.  Removing  $c_1$ or $c_2$, on the other hand, offers a ratio of $100\%$ for $p_1$ within the coalition, but the overall share of the coalition is $<50\%$.  Only eliminating $o_2$ strikes the right balance: the share of the entire coalition is $62\%$, and $p_1$ has $50\%$ of this, as required.

\input{example-table}


\section{Bribery}\label{sec:bribery}
Bribery allows the manipulator to change only the voters' preference orders. So, $\hat{P}=P$, and $\hat{V}=V$.  A modification of $\succ_i$ to $\hat{\succ}_i$ is associated with a \emph{cost} (which is positive only if $\succ_i \neq \hat{\succ}_i$), and the manipulator has a budget $B\geq 0$.

Technically, we equate a bribe with the resultant sequence of modified orders $\bbribe=(\sbribe_{v_1},\ldots,\sbribe_{v_n})$. 
The cost of changing voter $v_i$'s preference order from $\oi_{i}$ to $\sbribe_{i}$ is $\costIbribery{i}{\oi_{i}}{\sbribe_{i}}$. 
The cost of the entire bribe is $\costbribery{\order}{\bbribe}=\sum_{i=1}^n\costIbribery{i}{\oi_{i}}{\sbribe_{i}}$.
When clear from the context, we omit the original order $\oi_{i}$, writing $\costIbriberyshort{i}{\sbribe_{i}}$.
The sub-types of the problem are described below.

\paragraph{The Decision Problem}
The core decision problem is:






\begin{boxC}{gray}{\textbf{Bribery in Parliamentary Elections}}
\textbf{Input: }\hspace{0.1em} (P, V, $\runningorders$, B, $Cost$, $\mathcal{O}$) where:
\begin{tabular}{l l l}
& $P$: Parties, \quad $V$: Voters, \quad $\runningorders$: Preferences, \quad
$B$: Budget \\
& $Cost$: Cost function, \quad $\mathcal{O}$ : Objective (J or J+F)
\end{tabular}


\tcblower
\textbf{Output: } \begin{tabular}{l l}
&Does there exist a bribe $\bbribe$ with $\costbribery{\order}{\bbribe} \leq B$ such that the briber’s
objective $\mathcal{O}$ is met in the\\
&election $\tilde{E}:=(P, V, \bbribe)$? 
\end{tabular}
\end{boxC}


\subsection{Bribery Types}
The literature considers several different bribery types, distinguished both by the permissible changes to the voters' preferences and by the cost to do so. In this paper, we consider the following four bribery types:

\subsubsection{\textbf{$1$-Bribery:}} All changes to voters' preference orders are permissible. The cost for all voters and all changes is constant and identical; that is, $\costOneB{i}{\oi_{i}}{\sbribe_{i}}= 1$ for all $i,\oi_{i},\sbribe_{i}$.   
    
\subsubsection{\textbf{$\$$-Bribery:}}  All changes are permissible and the cost depends on the voter. That is, for each $v_i$, there is a constant value $\pi_i$, with $\costDolarB{i}{\oi_{i}}{\sbribe_{i}}= \pi_i$, for any $\oi_{i}$ and $\sbribe_{i}$~\cite{faliszewski2006complexity}.

\subsubsection{\textbf{Swap-Bribery:}} All changes are permissible.  The cost depends on pairs of parties that swapped order between the preference orders~\cite{elkind2009swap}. Specifically, for voter $v_i$ and parties $p,q$, there is a cost $sw_i(p,q)$ for moving $q$ from a location below $p$ to a location above $p$ in $v_i$'s order. The total cost for $\sbribe_i$ is the sum:
    $$\costSwapB{i}{\oi_{i}}{\sbribe_i}=\sum_{(p,q) ~:~ p \succ_{i} q \text{ and } q \sbribe_{i} p}sw_i(p,q)$$

\subsubsection{\textbf{Coalition-Shift-Bribery:}} In the single-winner setting, shift-bribery allows only to shift the preferred candidate and only upward. The cost is a monotone function of the number of locations this candidate is shifted~\cite{elkind2009swap}. 
    We adapt this definition to the coalitional setting.
    In \emph{coalition-shift-bribery} only the parties of the coalition $C$ can be shifted and only upward (a swap between two coalition parties is also permitted).  
    That is, if $p \succ_{i} q$ and $q \sbribe_{i} p$ then $q\in C$.  
    The cost is: 
    $$\costShiftB {i}{\oi_{i}}{\sbribe_{i}}
    :=s_i(|\{ (p,q) : p \succ_{i} q \text{ and } q \sbribe_{i} p \}|)$$
    for some monotone function $s_i$, depending on $i$'s.

\subsection{Bribery: Summary of Results}
The results are summarized in Table~\ref{tab:sum_complexity_bribery}.
For 1-Bribery and \$-Bribery all variants are polynomial-time solvable (\Cref{thm:dolar-Plurality-t-CBPP}).  
For Swap and Coalition-Shift-Bribery the complexity depends on the electoral threshold; when $\thrs=0$ the problems are polynomial-time solvable (\Cref{thm:Plurality-ccbp}); but the problems are NP-hard when $\thrs>0$ (\Cref{cor:swap-Plurality-t-CBP} and \Cref{thm:shift-Plurality-t-CBP}).


\begin{table}[h]
\centering
\small
\setlength{\tabcolsep}{3pt}
\begin{tabular}{|c|l|l|l|}
\hline
\multicolumn{2}{|c|}{\textbf{Bribery Type}} & {\textbf{J Objective}} & \textbf{J+F Objective}\\
\hhline{|==|=|=|}
\multicolumn{2}{|c|}{$ 1$}
    & P ~~~(Thm. \ref{thm:dolar-Plurality-t-CBPP}$^*$)
    & P ~~~(Thm. \ref{thm:dolar-Plurality-t-CBPP}$^*$) \\
\hhline{|==|=|=|}
\multicolumn{2}{|c|}{$\$$}
    & P ~~~(Thm. \ref{thm:dolar-Plurality-t-CBPP}$^*$)
    & P ~~~(Thm. \ref{thm:dolar-Plurality-t-CBPP}) \\
\hhline{|==|=|=|}
\multirow{2}{*}{Swap} & $\thrs=0$
    & P ~~~(Thm. \ref{thm:Plurality-ccbp}$^*$)
    & P ~~~(Thm. \ref{thm:Plurality-ccbp}) \\
\cline{2-4}
& $\thrs>0$
    & NP-hard ~(Thm. \ref{cor:swap-Plurality-t-CBP})
    & NP-hard ~(Thm. \ref{cor:swap-Plurality-t-CBP}$^*$) \\
\hhline{|==|=|=|}
Coalition & $\thrs=0$
    & P ~~~(Thm. \ref{thm:Plurality-ccbp}$^*$)
    & P ~~~(Thm. \ref{thm:Plurality-ccbp}) \\
\cline{2-4}
Shift & $\thrs>0$
    & NP-hard ~(Thm. \ref{thm:shift-Plurality-t-CBP})
    & NP-hard ~(Thm. \ref{thm:shift-Plurality-t-CBP}$^*$) \\
\hline
\end{tabular}
\caption{Bribery: Summary of complexity results. In parentheses --- the corresponding theorem. An astrict denotes that the result \emph{follows} directly from the theorem.}
\label{tab:sum_complexity_bribery}
\end{table}

\subsection{1-Bribery and \$-Bribery}\label{sec:brib-1-and-dolar}

\begin{toappendix}
    \subsection{1-Bribery and \$-Bribery}
\end{toappendix}

In this section, we show that, for 1-bribery and \$-bribery, the problem can be solved in polynomial time even when the briber's objective is J+F. 
Notice that this implies that all other variants are polynomial-time solvable as well.


\begin{theoremrep}
     \problem{B}{\$}{J+F}{\geq}  is polynomial-time solvable.
     \label{thm:dolar-Plurality-t-CBPP}
\end{theoremrep}

The complete proof is given in the appendix (see \Cref{alg:dolar-Plurality-t-CBPP}).

\begin{proofsketch}
    Since in 1‑bribery and \$‑bribery any change is permissible and the cost per voter is constant, the problem can be approached in three steps: (1) iterate over every set of voters that could be bribed; (2) compute the optimal bribe for that set; and (3) check whether that bribe achieves the briber’s objectives. 
    
    We show two things:    
    first, that using a dynamic‑programming algorithm we can efficiently consider all relevant sets of voters that could be bribed; and second, that for each such set, the optimal bribery can be computed in polynomial time. If at least one feasible bribe achieves the briber’s objectives, we assert that the answer is “yes”; otherwise, it is "no".
\end{proofsketch}

\begin{proof}
To prove the theorem, we show that  Algorithm \ref{alg:dolar-Plurality-t-CBPP} (described below) solves the problem and runs in polynomial time.
The algorithm splits the bribing process into two: first, determining the set of voters to bribe --- and omitting their current votes from the pool of votes --- and only afterward determining their new votes.   
Accordingly, we define the notion of an \textit{undetermined bribe}, which is the situation wherein a set of voters is bribed, thus eliminating their original votes, but their new votes are not yet determined.  
Technically, an undetermined-bribe is a set $\tilde{V} =\{ v_{i_1},\ldots,v_{i_{\ell}}\}$ of (bribed) voters. 
 The cost of the undetermined bribe $\tilde{V}$ is $\costDolarBSet{\tilde{V}}=\sum_{i:v_i\in \tilde{V}}\pi_i$. 
 Note that under \$-bribery this cost suffices to later determine the votes of all members of $\tilde{V}$ in any way one may wish. 

The key ingredient in the algorithm is a dynamic programming computation of a function $g$, such that $g(\ell, a_{\bar{C}},d,a_{C_{-1}})$ is the least cost of an undetermined-bribe $\tilde{V}$ that satisfies all the following four requirements. (1) $|\tilde{V}| = \ell$, (2) $\topT{E}{v_i}\neq p_1$, for all $v_i\in \tilde{V}$, (3) from the votes of $V\setminus \tilde{V}$ (the not bribed voters), there are $a_{\bar{C}}$ active votes for parties of $\bar{C}$; and (4) there is a way to distribute $d$ additional votes to parties of $C_{-1}$ in a way that, together with the votes of $V\setminus \tilde{V}$, the total number of active votes for parties of $C_{-1}$ is $a_{C_{-1}}$.  
If there is no undetermined bribe that satisfies the above, then $g(\ell, a_{\bar{C}},d,a_{C_{-1}})=\infty$.

Let $mincost(p_j,\ell)$ be the least cost of bribing $\ell$ voters who voted for $p_j$ in $E$ (with $\infty$ if there is no such bribe).  

We shall now prove that it is possible to compute $g(\ell, a_{\bar{C}},d, a_{C_{-1}})$, for all $0\leq \ell, a_{C_{-1}}, d, a_{\bar{C}}\leq n$, and $mincost(p_j,k)$, for $1\leq j\leq m, 1\leq \ell \leq n$, in polynomial time.

    For party $p_j$, let $V_j=\{ v_i : \topT{E}{v_i}=p_j\}$ --- the set of voters who (before bribes) vote for $p_j$.  For a set of parties $D$, set $V_D=\cup_{p_j\in D}V_j$. 
    For computing $mincost(p_j,\ell)$ it suffices to first sort the set $\{ \pi_i : v_i\in V_j\}$.  Then $mincost(p_j,\ell)$ is the sum of the $\ell$ least values in this set.
    
    For $g$:
    When seeking the least-cost undetermined bribe $\tilde{V}$, we separately compute $\tilde{V}_{C_{-1}}$ --- the bribed voters who initially voted for $C_{-1}$ --- and $\tilde{V}_{\bar{C}}$ --- the bribed voters who initially voted for $\bar{C}$.
    
    \paragraph{Computing $\tilde{V}_{\bar{C}}$}
    For a set of parties $D\subseteq \bar{C}$, let $f(D,\ell,a_D)$ be the least cost of an undetermined bribe $\tilde{W}\subseteq V_D$ such that:  (i) $|\tilde{W}|=\ell$, (ii) based on the votes of $V_D\setminus \tilde{W}$ exactly $a_D$ votes are active.  
    
    We provide a dynamic programming process for computing $f(D,\ell,a_{\bar{C}})$ for all $D\subseteq \bar{C}$, and $0\leq \ell,a_{\bar{C}}\leq n$.  First, for $D$ of size 1:
    \begin{align}
        f(\{j\},\ell,a_{\{ j\}})= mincost(p_j,\ell) \nonumber
    \end{align}
    if either $a_{\{j\}}=|V_j|-\ell\geq T$ or $|V_j|-\ell< T$ and $a_{\{j\}}=0$.  In all other cases, $f(\{j\},\ell,a_{\{ j\}})=\infty$. 
    
    Now, 
    \begin{align*}
        f(D\cup \{j\},\ell,a_{D\cup\{j\}})=
    \;\; \min_{0\leq \ell'\leq \ell, 0\leq a_{D} \leq a_{D\cup \{ j\}}} 
    \{     f(D,\ell',a_D) +     f(\{ j\},\ell-\ell',a_{D\cup\{ j\}}-a_D) \} 
    \end{align*}
    This allows to iteratively compute $f(\bar{C},\ell,a_{\bar{C}})$, for all $\ell,a_{\bar{C}}$, by adding parties one by one.

    \textbf{Computing $\tilde{V}_{C_{-1}}$.}
    For a set of parties  $D\subseteq C_{-1}$, let $h(D,\ell,d , a_{D})$ be the least cost of an undetermined bribe $\tilde{W}\subseteq V_D$ such that:  (i) $|\tilde{W}|=\ell$, (ii) with $d$ additional votes, it is possible to obtain a total of $a_{D}$ active votes for $D$.
    

    Again, we use dynamic programming to compute
    $h(C_{-1},\ell,d,a_{C_{-1}})$ for all $0\leq \ell,d, a_{\bar{C}}\leq n$.
    
    For $D$ of size 1, 
    if either one of the following holds: 
        \begin{itemize}
            \item $|V_j|-\ell+d\geq T$ and $a_{\{ j\}}=|V_j|-\ell+d$,
            \item $|V_j|-\ell+d< T$ and $a_{\{ j\}}=0$
        \end{itemize} 
        then we set $h(\{j\},\ell,d,a_{\{ j\}})$ to $mincost(p_j,\ell)$; otherwise, we set it to $\infty$. Now, 
    \begin{align*}
        h(D\cup \{j\},\ell,d,a_{D\cup\{j\}})=
    \;\; \min_{0\leq \ell'\leq \ell, 0\leq d'\leq d, 0\leq a_{D} \leq a_{D\cup\{ j\}}} \{     h(D,\ell',d',a_D) +  h(\{ j\},\ell-\ell',d-d',a_{D\cup\{ j\}}-a_D) \} 
    \end{align*}
    This allows to iteratively compute $h(C_{-1},\ell,d,a_{C_{-1}})$, for all $\ell,d,a_{C_{-1}}$, by adding parties one by one.
    
    \paragraph{Computing $g$}
    Now, having computed $h(C_{-1},\ell,d,a_{C_{-1}})$ and $f(\bar{C},\ell,a_{\bar{C}})$ for all $\ell, d,a_{C_{-1}},a_{\bar{C}}$, we compute $g$:
    \begin{align*}
        g(\ell,a_{\bar{C}},d,a_{C_{-1}}) =& 
    \;\; \min_{0\leq \ell'\leq \ell} \{  h(C_{-1},\ell',d,a_{C_{-1}})
    +f(\bar{C},\ell-\ell',a_{\bar{C}})\}  \;\;\;\;\;\;\;\; \square
    \end{align*} 


Then, given the efficient computation of $g$, Algorithm \ref{alg:dolar-Plurality-t-CBPP} 
iterates over all possible combinations of $\ell, a_{\bar{C}},d, a_{C_{-1}}$, and for each, performing three steps: First, we compute $g(\ell, a_{\bar{C}},d, a_{C_{-1}})$ and check if the budget suffices (line 2). 
\hspace{0.3em} If so, then we try to add $d$ bribed votes to support $C_{-1}$ and the rest to $p_1$ (lines 2-6).
The bribe of $g$ provides $\ell$ ``free'' votes, while we are seeking to add $d$ votes for $C_{-1}$.  If $d>\ell$, then we need to add $d-\ell$ votes from $p_1$.  Line 3 check if this is possible. Line 4 adds the remaining $\ell-d$ (which may be negative) to the support of $p_1$. 
\hspace{0.3em} Next, we check if the bribe is successful (lines 5-7):
Lines 5-6 adjust the seats of $p_1$ according to the threshold. Finally, line 7 checks whether the bribe satisfies the required target support and target ratio ($\varphi$ and $\rho$). 
If so, in line 8, we assert that the budget is sufficient. Otherwise, the algorithm continues to the next iteration.
If no valid bribe is found in any iteration, the algorithm asserts that the budget is insufficient (line 9). 

\end{proof}

\begin{toappendix}
    
\begin{algorithm}
    \begin{algorithmic}[1]
        \FORALL{$0\leq \ell, a_{\bar{C}},d, a_{C_{-1}} \leq n$}        
            \IF{$g(\ell, a_{\bar{C}},d, a_{C_{-1}}) \leq B$}
            \IF{$d\leq \ell$ \textbf{and} $mincost(p_1, d-\ell) \leq B-B'$}
            \STATE $a_{p_1}=\Npoint{E}{p_1}+\ell-d$
            \IF{($a_{p_1} < T$) }
                \STATE $a_{p_1}=0$
            \ENDIF
            \IF{$\frac{a_{C_{-1}}+a(p_1)}{a_{C_{-1}}+a(p_1) + a_{\bar{C}}} \geq \varphi \text{ and }  \frac{a(p_1)}{a_{C_{-1}}+a(p_1)}\geq \rho$}
                \RETURN Budget is \emph{Sufficient}.
            \ENDIF
            \ENDIF
            \ENDIF
        \ENDFOR
    \RETURN Budget is \emph{Insufficient}.
    \end{algorithmic}
    \caption{Algorithm for \$-Bribery for J+F-ob}
    \label{alg:dolar-Plurality-t-CBPP}
\end{algorithm}

\end{toappendix}




\subsection{Swap and Coalition-Shift Bribery, $\thrs=0$}\label{sec:brib-swap-shift-no-threshold}

\begin{toappendix}
    \subsection{Swap and Coalition-Shift Bribery, $\thrs=0$}\label{sec:brib-swap-shift-no-threshold}
\end{toappendix}

With no electoral threshold, Swap and Coalition-Shift Bribery are polynomial-time solvable even when the briber's objective is J+F.


\begin{theoremrep}
\problem{B}{Swap}{J+F}{=} and \problem{B}{Coalition-shift}{J+F}{=} are polynomial-time solvable.
    \label{thm:Plurality-ccbp}
\end{theoremrep}

The complete proof is given in the appendix.
\begin{proofsketch}
    
We show that the problem can be decided by solving $O(n^2)$ instances of the \emph{minimum-cost flow (MCF)} problem, which is (loosely) defined as follows: given a flow network with a cost and capacity on each edge, and a demand $d$, the goal is to find the cheapest feasible way to send $d$ units of flow from the source $s$ to the target $t$. MCF can be solved in polynomial time (see e.g.~\cite{ahuja1993network}).

The idea is to check all possible divisions of votes among the favored party $p_1$, the coalition without it $C_{-1}$, and the opposition $\bar{C}$ that satisfy the $J+F$ objective, and check whether the briber can achieve such a division by solving an MCF instance as follows.

The flow network is described in Figure 1. The flow represents the number of voters.
The nodes on the second layer (the one after the source) represent the number of votes that go to $p_1$, $C_{-1}$, and $\bar{C}$. The cost of the edges from $s$ to these nodes is 0, and their capacities equal the number of voters that \emph{need} to vote for each group in this division (so if these edges are saturated, we obtain exactly the desired division of votes).

The nodes on the third layer represent the voters’ true preference orders. Each edge from the second layer to the third has cost 0 and capacity 1 (hence each unit of flow corresponds to a specific voter voting for a specific group).

The nodes on the fourth layer represent the votes' preferences after the bribery, and each edge from the third layer to the fourth represents the cost of changing this particular voter’s preference in the required way. The cost is determined by the bribery cost function, and the capacity is 1.

Finally, each node in the fourth layer is connected to the target $t$ with an edge of cost 0 and capacity 1.

The total demand is $n$, corresponding to all voters.
If the minimum cost of the resulting flow is at most the briber’s budget $B$, then the desired outcome is achievable; otherwise, it is not.
\end{proofsketch}

\begin{figure}[tbhp]
    \centering
    \includegraphics[width=0.7\linewidth]{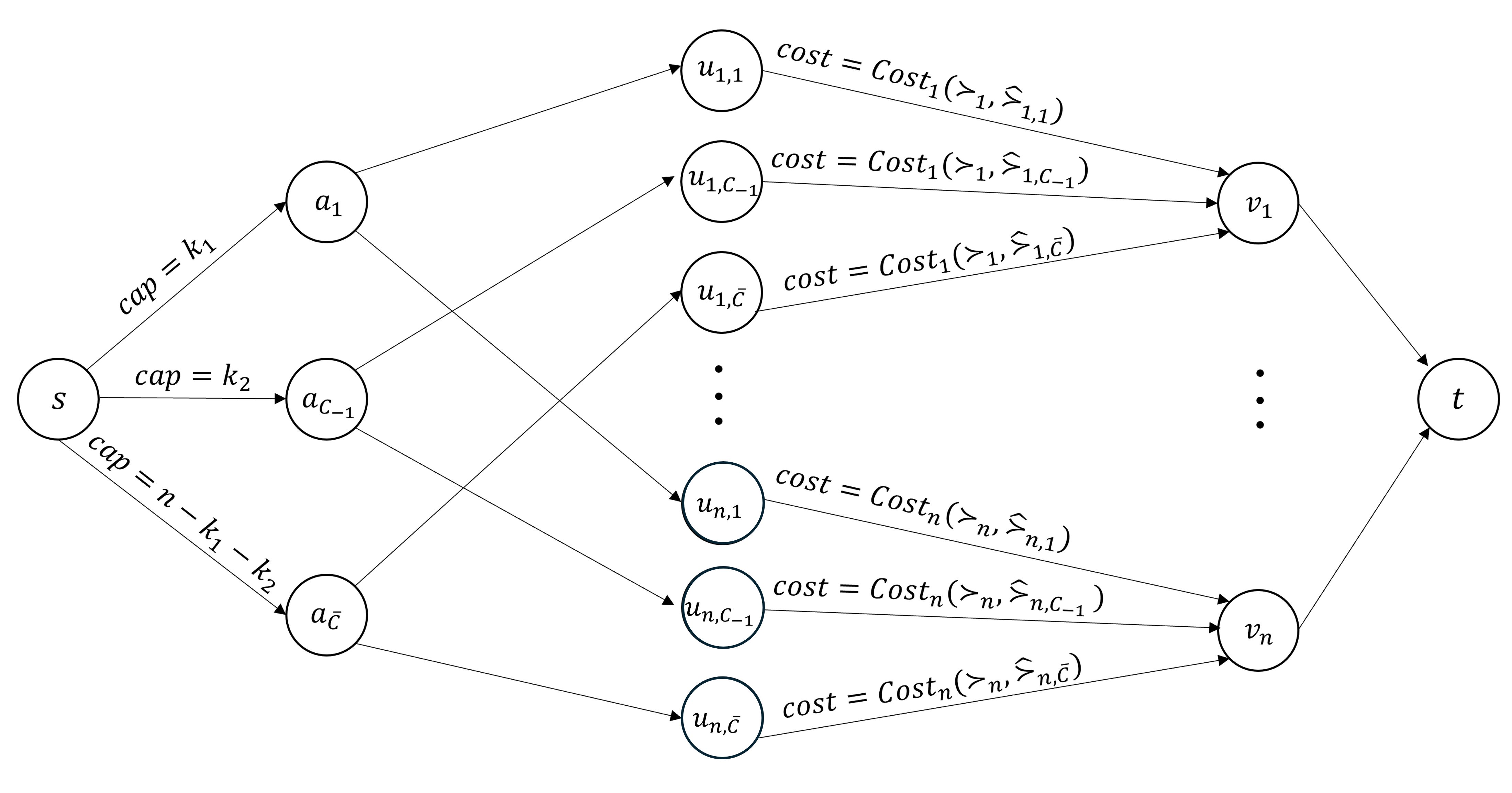}
    \caption{Network for proof of Theorem \ref{thm:Plurality-ccbp}. Unless otherwise stated, the default values are $cost = 0$ and $cap = 1$. $cap$ stands for capacity.} 
   
    \label{fig:Plurality-ccbp}
\end{figure}

\begin{proof}
    
We show that the problem can be decided
 by solving $O(n^2)$ 
instances of the minimum-cost flow (MCF) problem, defined as follows.

\begin{definition}[Minimum-Cost Flow (MCF)]
    Given: a directed graph $G=(U, D)$, source and sink nodes $s,t\in U$, 
    non-negative capacity and cost functions $cap(e)$, $cost(e)$ for each $e \in D$, and demand $d$.
    The goal is to find a flow $f: D \rightarrow \mathbb{R}$  that minimizes the weighted cost $\sum_{e\in D}f(e)cost(e)$, while satisfying: (i) capacity constraint:  $0\leq f(e)\leq cap(e)$ for all $e\in D$, (ii) flow conservation: 
    $\sum_{u\in V}f(u,v)= \sum_{u\in V}f(v,u)$, for $v\in V\setminus \{ s,t\}$, (iii) demand fulfilled: $\sum_{u\in V}f(s,u)= \sum_{u\in V}f(u,t)=d$.  
\end{definition}

MCF can be solved in polynomial time (see e.g.~\cite{ahuja1993network}).
We show how, given $k_1,k_{C_{-1}}$, to construct an MCF instance $M_{k_1,k_{C_{-1}}}$ that admits a solution with weighted cost
 $B'$ iff there exists a bribe $\bbribe$ with cost $B'$ such that $\Npoint{\hat{E}}{p_1}=k_1$ and $\Npoint{\hat{E}}{C_{-1}}=k_{C_{-1}}$. 

For each $i$, let $\sbribe_{i,p_1}$ be the bribe of $v_i$ that brings $p_1$ to the top position (without moving any other parties).  Similarly, let $\sbribe_{i,C_{-1}}$ 
and $\sbribe_{i,\bar{C}}$ are the minimal cost bribes of $v_i$ that bring a member of $C_{-1}$ and of $\bar{C}$ to the first position, respectively.  Note that $\sbribe_{i,\cdot}$ may be $\oi_i$, if no bribe is necessary. Also, for determining $\sbribe_{i,C_{-1}}$ and $\sbribe_{i,\bar{C}}$ we only need to consider bribes that bring a single party to the top position, with no other changes. So, all these bribes, together with their costs, can be computed in polynomial time for both Swap-Bribery and Coalition-Shift-Bribery.

The graph, capacities, and costs in $M_{k_1,k_2}$ are defined as follows (see Figure \ref{fig:Plurality-ccbp}).  The graph nodes are:
One source node $s$ and one target node $t$, three nodes $a_{1},a_{C_{-1}}$ and $a_{\bar{C}}$; and for each voter $v_i$ four nodes: $v_i,u_{i,1}, u_{i,C_{-1}}$ and $u_{i,\bar{C}}$.
The edges, capacities, costs, and demand are:
\begin{itemize}
    \item edges $(s,a_{1}), (s,a_{C_{-1}}), (s,a_{\bar{C}})$, all with cost 0 and capacities  $cap(s,a_{1})=k_1,cap(s,a_{C_{-1}})=k_2,cap(s,a_{\bar{C}})=n-k_1-k_2$.  
    \item for each $i$, edges $(a_{1},u_{i,1}), (a_{C_{-1}},u_{i,C_{-1}})$ and $(a_{\bar{C}},u_{i,\bar{C}})$, all with capacity $1$ and cost $0$.
    \item for each $i$,  edges $(u_{i,\alpha},v_i)$ for $\alpha\in \{ 1,C_{-1},\bar{C}\}$, with $cost(u_{i,\alpha},v_i)= \costIbribery{i}{\universeOi_i}{\sbribe_{i,\alpha}}$, and capacity 1.
    \item for each $i$,  edge $(v_i,t)$ with cost 0 and capacity 1.
    \item the demand is $d=n$.
\end{itemize}

We shall now prove that 
    $M_{k_1,k_{C_{-1}}}$  admits a solution with weighted cost $B'$ iff there exists a bribe $\bbribe$ with cost $B'$ such that $\Npoint{\influencedE}{p_1}=k_1$ and $\Npoint{\influencedE}{C_{-1}}=k_{C_{-1}}$.
    
Let $f$ be a flow that solves $M_{k_1,k_{C_{-1}}}$ with cost $B'$.  We construct a bribe $\bbribe$ with the same costs as claimed by the lemma. Since all inputs are integral we may assume that $f$ is integral.  Since the demand is fulfilled, and $cap(v_i,t)=1$, it must be that $f(v_i,t)=1$ for all $i$.  Hence, for each $i$, there exists exactly one $\alpha_i\in \{ 1,C_{-1},\bar{C} \}$ with $f(u_{i,\alpha_i},v_i)=1$.  So, define $\sbribe_i=\sbribe_{i,\alpha_i}$. By construction, the cost of this bribe is $\costIbribery{i}{\universeOi_i}{\sbribe_{i,\alpha_i}}$, and the total cost of the entire bribe is $\sum_{i=1}^{n}\costIbribery{i}{\universeOi_i}{\sbribe_{i,\alpha_i}}$ --- which is also the weighted cost the flow $f$. Now, since the demand $d=n$ is fulfilled, $f(s,a_1)=k_1$ and $f(s,C_{-1})=k_{C_{-1}}$. So, there are exactly $k_1$ (res. $k_{C_{-1}}$) indexes $i$, with $f(a_{1},u_{i,1})=1$ (res. $f(a_{C_{-1}},u_{i,C_{-1}})=1$).  So, $\Npoint{\influencedE}{p_1}=k_1$ and $\Npoint{\influencedE}{C_{-1}}=k_{C_{-1}}$.

The proof of the other direction is similar.


Next, for each $0\leq k_1, k_{C_{-1}}\leq n$, $k_1+k_{C_{-1}} \leq n$, such that $\frac{k_1+k_2}{n} \geq \varphi$ and $\frac{k_1}{k_1+k_2} \geq \rho$, construct the MFC instance $M_{k_1,k_2}$.  Solve $M_{k_1,k_2}$.  If the solution is with cost $\leq B$, then, by the above claim, there is also a bribe $\bbribe$ with the same cost, which solves the associated bribery C+F-op instance, when $\thrs=0$. Hence, return ``success''. If none of the $M_{k_1,k_{C_{-1}}}$ admit a solution with cost $\leq B$, then assert "no solution".
The number of pairs $k_1,k_{C_{-1}}$ to check is $O(n^2)$. 
\end{proof}

\begin{toappendix}
    \begin{center}
    \includegraphics[width=0.7\linewidth]{media/shit-bcfp-2.png}
    
    {\small Figure~\ref{fig:Plurality-ccbp}: Network for proof of Theorem \ref{thm:Plurality-ccbp} (reproduced for convenience in the appendix). Unless otherwise stated, the default values are $cost = 0$ and $cap = 1$. $cap$ stands for capacity.}
\end{center}
\end{toappendix}

\subsection{Swap and Coalition-Shift-Bribery: $\thrs>0$}\label{sec:brib-swap-shift-with-threshold}

In this section we prove that when $\thrs>0$, all variants of Swap-Bribery and Coalition-Shift-Bribery are NP-hard.


    

\begin{theorem} \label{thm:shift-Plurality-t-CBP}
      \problem{B}{Coalition-shift}{J}{\geq} is NP-hard, even if when all $s_i(x)$'s are multiplicative.
\end{theorem}
This implies that also \problem{B}{Coalition-shift}{J+F}{\geq} is NP-hard.
The proof is by reduction from the 3-4-Exact-Cover.
\begin{proof}
    The proof is by reduction from the 3-4-Exact-Cover problem: Given a set $Z$, and a collection $\mathcal{D}$ of $4$-subsets of $Z$ such that any $z \in Z$ appears exactly in $3$ subsets in $\mathcal{D}$; it determines whether there is a sub-collection of $\mathcal{D}$ such that any element in $Z$ appears exactly once.

    We show that given an instance to the 3-4-Exact-Cover Problem, we can construct an instance of the \problem{B}{Coalition-shift}{J}{\geq} problem such that a cover exists if and only if a successful bribe exists. 
    
    The instance $E=(\universe, \runningVoters, \universeorders)$ is defined as follows. The set of parties $\universe$ contains one party $p_i$ for for any subset $D_i \in \mathcal{D}$; and $3|Z|$ additional parties:     
    $$\universe :=\{p_1, p_2,\ldots p_{|\mathcal{D}|}, ~p_{|\mathcal{D}|+1},\ldots p_{(|\mathcal{D}| + 3 |Z| +1)}\}$$

    There are $2|Z|$ voters, 2 voters corresponding to each element $z_i \in Z$, with the following preferences:
    \begin{itemize}
        \item $1$ voter of Type 1:  $p_{|\mathcal{D}|+1}$ in the first place, followed by the $3$ parties corresponding to the $3$ subsets that contain $z_i$, then the parties $p_j$ for $|\mathcal{D}|+2 \leq j \leq |\mathcal{D}|+3|Z| +1$, and then all other parties. 
        
        \item $1$ voter of Type 2: party $p_{|\mathcal{D}|+1}$ in the first place, followed by the parties $p_j$ for $|\mathcal{D}|+2 \leq j \leq |\mathcal{D}|+3|Z| +1$, and then all other parties. 
    \end{itemize}

    Let the coalition be $C :=\{p_1,\ldots p_{|\mathcal{D}|}\}$ and the favorite party be $p_1$.
    
    Recall that the bribery type is coalition-shift. Let the price function of each voter be $s_i(x) := x$; and the budget be $B:=3|Z|$.
    
    The target support is $\varphi := \frac{1}{2}$, and the election threshold is $\thrs:=\frac{4}{2|Z|}$.

    ($\Rightarrow$) We assume that there is an exact cover $\mathcal{D}$ and prove that there is a successful bribe.
    We prove that bribing all voters of Type 1, to vote for the party corresponding to the subset in $\mathcal{D}'$ that covers the corresponding element $z_i$ is feasible and successful. 

    First, the bribe is feasible as the cost of bribing each of these voters is at most $3$ (the parties corresponding to the sets contain the corresponding element are at the top $4$ places); the bribe involves $|Z|$ voters, and the budget is $3|Z|$.

    Second, the bribe is successful as afterwards (1) the opposition party $p_{|\mathcal{D}| +1}$ gets all the votes of voters of type 2 ($|Z|$ votes),  and since $\frac{1}{2} \geq \frac{2}{|Z|}$ all of these votes are active; (2) each coalition party correspond to a subset in $\mathcal{D}'$ gets exactly $4$ votes from the $4$ voters of type $1$ corresponding to the element in the subset, since $\thrs:=\frac{4}{2|Z|}$ and there are $2|Z|$ voters, $4$ is sufficient and all those votes are active; and (3) all other parties do not receive votes at all. 
     The coalition gets at least $1/2$ of the active votes, which means that the coalition objective is satisfied for $\varphi = \frac{1}{2}$.

    ($\Leftarrow$) We assume that there is a successful bribe and prove that there is an exact cover $\mathcal{D}$.

    Consider any $p_i \in C$. 
    Before the bribe, it does not get any votes. 
    It is impossible to bribe the voters of type 2 to vote for the coalition as there are $3|Z|+1$ parties before them so each such change will cost more than the budget $3|Z|$. 
    Notice that this shift bribe only allows us to promote the coalition parties. Thus, there are $|Z|$ votes for $p_{|\mathcal{D}|+1}$ (from all voters of type 2).

    In order to have at least $1/2$ of the active votes for the coalition, there must be $|Z|$ active votes for $c \in C$.
    As there are $2|Z|$ voters and $\thrs=\frac{4}{2|Z|}$, only parties that receive at least $4$ votes pass the election threshold.
    This means that a party $p_i$ will be part of the parliament only if all $4$ voters of Type 1 corresponding to the elements of the set $D_i$ change their preference order to rank $p_i$ in the first place.
    Notice that it is possible only to bribe a voter $v_i$ of Type 1 to vote for the coalition parties that corresponds to a subset that contains $z_i$ (since there are $3|Z|+4$ parties before the other parties from $C$).
    
    In order to satisfy $\sum_{c \in C} (\seats{\thrs}{\influencedE}(c)) \geq\frac{1}{2}$, there must be $\frac{n}{4}$ parties from $P$ that receive $4$ votes.  
    These $\frac{n}{4}$ parties correspond to $\frac{n}{4}$ subsets that form an exact cover.
\end{proof}

\begin{theorem}
    \problem{B}{Swap}{J}{\geq} is NP-hard.
    \label{cor:swap-Plurality-t-CBP}
\end{theorem}
\begin{proof}
Given an instance of coalition shift bribery with multiplicative $s_i$'s, we reduce it to an instance of swap bribery. For voter $v_i$ and $x,y\in \universe$ define $sw_i(x,y)=(s_i)'$ for $y\in C$ (where $(s_i)'$ is the slope of $s_i$), and $sw_i(x,y)=B+1$ otherwise.  The latter provides that only allowed swaps will be considered, and the former that the cost remains the same.
\end{proof}



\section{Voter Control}\label{sec:voter-control}

Voter-Control allows the chair to change only the set of voters.
Here, the set of parties remains unchanged, $\hat{P} := P$, while the set of voters is modified to $\hat{V}$ (as described below). The preferences are $\hat{\succ}:=(\succ_{v})_{v\in \hat{V}}$ --- they are adapted to reflect only the voters in $\hat{V}$ (the manipulator \emph{cannot} change the voters' preferences). The chair has a budget $B \ge 0$, and each change incurs a \emph{cost}.

\subsection{Voter Control Types}

We consider two types of voter control. The first is control by deleting voters (DV), in which the chair can choose a subset of voters to remove from the election. 
The second is control by adding voters (AV), where the chair can introduce additional voters with predetermined preferences.

\subsubsection{\textbf{Control by Deleting Voters (DV)}}
The manipulator can choose to eliminate a subset $V^{-}\subseteq V$ of voters. 
Deleting voter $v \in \universeVoters$ incurs a cost $\costVoterControl{v}$. For a subset of voters, the total cost is the sum of the costs of individual deletion.


\begin{boxC}{gray}{\textbf{Control by Deleting Voters in Parliamentary Elections}}
\textbf{Input: }\hspace{0.1em} (P, V, $\runningorders$, B, $Cost$, $\mathcal{O}$) where:
\begin{tabular}{l l l}
& $P$: Parties, \quad $V$: Voters, \quad $\runningorders$: Preferences, \quad
$B$: Budget \\
&$Cost$: Cost function, \quad $\mathcal{O}$ : Objective (J or J+F)
\end{tabular}

\tcblower
\textbf{Output: } 
\begin{tabular}{l l}
&Does there exist a set $V^- \subseteq \universeVoters$ with $\costVoterControl{V^-} \leq B$ such that the chair's objective $\mathcal{O}$ is met in the\\
&election $\tilde{E}:=(P, \hat{V}, \bbribe)$ where $\controlV := \universeVoters \setminus V^-$ and $\bbribe:=(\succ_{v})_{v\in \hat{V}}$?
\end{tabular}
\end{boxC}

\subsubsection{\textbf{Control by Adding Voters (AV)}}
We are given a set of \emph{spoiler} voters, $\spoilerVoters$,  $\spoilerVoters \cap  \universeVoters=\emptyset$, each associated with a preference order.
The chair may choose to add a subset $W^{+}\subseteq W$ of spoiler voters.  
Adding a voter $v \in \spoilerVoters$ incurs a cost $\costVoterControl{v}$. For a subset of voters, the total cost is the sum of individual additions.



\begin{boxC}{gray}{\textbf{Control by Adding Voters in Parliamentary Elections}}
\textbf{Input: }\hspace{0.1em} (P, V, W, $\runningorders$, B, $Cost$, $\mathcal{O}$) where:
\begin{tabular}{l l l}
& $P$: Parties, \quad $V$: Voters, \quad $W$: Spoiler Voters, \\ & $\runningorders$: Preferences of all voters (in $V\cup W$), \quad
$B$: Budget, \\
&$Cost$: Cost function, \quad $\mathcal{O}$ : Objective (J or J+F)
\end{tabular}

\tcblower
\textbf{Output: } 
\begin{tabular}{l l}
&Does there exist a set $W^+ \subseteq \spoilerVoters$ with $\costVoterControl{W^+} \leq B$ such that the chair's objective $\mathcal{O}$ is met in the\\
&election $\tilde{E}:=(P, \hat{V}, \bbribe)$ where $\controlV := V \cup W^+$ and $\bbribe:=(\succ_{v})_{v\in \hat{V}}$?

\end{tabular}
\end{boxC}

\subsection{Polynomial-Time Algorithm for Voter Control} 
\Cref{thm:voter-control} shows both types of voter control --- by deleting voters and by adding voters --- are polynomial-time solvable by extending the algorithms of \citet{put2016complexity}.



\begin{theoremrep}\label{thm:voter-control}
    \problem{C}{AV}{J+F}{\geq} and \problem{C}{DV}{J+F}{\geq} are polynomial-time~solvable.
    \label{thm:plurality-t-VCCFP}
\end{theoremrep}

The complete proof is given in the appendix.

\begin{proofsketch}
    We provide the proof for adding voters here; in the appendix, we show that deleting voters can be solved using the same technique via a relatively simple reduction.
    
    We consider all possible numbers of voters that could be added $r\in \{ 0\ldots |W|\}$, and all possible ways to divide $r$ into three $(a_1, a_{C_{-1}}, a_{\bar{C}})$: $a_1$ represents the number of voters we want to add to the preferred party $p_1$; $a_{C_{-1}}$ represents the number of voters we want to add to the coalition without it $C_{-1}:=C\setminus \{p_1\}$; while $a_{\bar{C}}$ represents the number of voters we want to add to the opposition $\bar{C}:=P \setminus C$.

    We prove that we can use a dynamic‑programming algorithm to efficiently iterate over all possible values of $r$ and divisions of $r$: $(a_1, a_{C_{-1}}, a_{\bar{C}})$.
    For each division, we can directly verify whether the chair’s objective is satisfied.
    If it does, we check if it is possible to add voters in this way: we prove that we can calculate, in polynomial time, what is the minimum cost required to achieve this additional support (if it cannot be achieved, we set the cost to $\infty$). 
    The key observation is that the relevant potential voters for each set—the preferred party ${p_1}$, the coalition without it $C_{-1}$, and the opposition $\bar{C}$—are distinct, because the sets themselves are disjoint.
    Therefore, for each group we only consider its own pool of relevant voters, sorted in non-decreasing order of cost.
    
    If the minimum cost of some division that satisfies the chair’s objective, does not exceeds the budget, then we assert that the answer is "yes"; otherwise, we assert "no." 
    
\end{proofsketch}

\begin{proof}
We start with the case of adding voters.    
    The algorithm iterates over all possible values of $r\in \{ 0\ldots |W|\}$, and for each determines if it is possible to reach the desired objectives by adding exactly $r$ voters.   

We first describe the process for a specific $r$. Set $T_r=\lceil \thrs\cdot (n+r)\rceil$ --- the electoral threshold with $r$ additional voters.  Say that a party is \emph{$T_r$-active} if its number of votes is $\geq T_r$.
Let $cost(p_i, s)$ be the least cost of adding $s$ voters from $W$ that vote to $p_i$. If there are no such $s$ voters, or $s<0$, then $cost(p_i, s) = \infty$. By sorting the voters of $W$ that vote for $p_i$ by cost, $cost(p_i, s)$ can be computed in polynomial time. 

Let $g(r,a_1,a_{C_{-1}},a_{\bar{C}})$ be the least cost of a set $W^+\subseteq W$ such that:  (1) $|W^+|=r$, (2) for $\hat{E}=(P,\hat{V},\hat{\succ})$, where $\hat{V}=V\cup W^+$, $\hat{\succ}=( \succ_v)_{v\in \hat{V}}$, it holds that the number of active votes to $p_1$ is $a_1$ to $C_{-1}$ is $a_{C_{-1}}$, and to $\bar{C}$ is $a_{\bar{C}})$ 

We shall now prove that it is possible to compute  $g(r,a_1,a_{C_{-1}},a_{\bar{C}})$, for any $r=0, 
    \ldots, |W|$ and $a_1, a_{C_{-1}}, a_{\bar{C}} \in [n+r]$, in polynomial time. 
For this proof, let $f_r(i,\ell, \bar{\ell},s)$ be the least cost of a set $W^+\subseteq W$ such that:  (1) $|W^+|=s$, (2) voters of $W^+$ vote for parties of $\{p_2,\ldots, p_i\}$, (3) based on the votes of $V\cup \hat{W}$ there are $\ell$ $T_r$-active votes to $\{ p_2,\ldots,p_i\}\cap C_{-1}$, and $\bar{\ell}$ $T_r$-active votes to $\{ p_2,\ldots,p_i\}\cap \bar{C}$. 

    Then, $f(i,\ell,\bar{\ell}, s)$ can be computed using dynamic programming, as follows.  For $i=1$, $f(1,\ell,s)=0$ if $\ell =s=0$ and  $f(1,\ell,\bar{\ell},s)=\infty$, otherwise.
    For $i=2,\ldots |C|$:
    \begin{multline*}
    f_r(i,\ell, \bar{\ell},s) = 
    \min\Big\{ f_r(i-1,\ell,\bar{\ell},s), \min_{\substack{
        T_r-N^E(p_i) \leq s' \leq s}}
        (cost(i,s') + f(i-1, \ell-(N^E(p_i)+s'),\bar{\ell}, s-s')) \Big\} 
    \end{multline*}
    For $i > |C|$: 
       \begin{multline*}
    f_r(i,\ell, \bar{\ell},s) = 
    \min\Big\{ f_r(i-1,\ell,\bar{\ell},s), \min_{\substack{
        T_r-N^E(p_i) \leq s' \leq s}}
        (cost(i,s') + f(i-1, \ell,\bar{\ell} -(N^E(p_i)+s'), s-s')) \Big\} 
    \end{multline*}
    
    Then,
    \begin{multline*}
    g(r,a_1,a_{C_{-1}},a_{\bar{C}}) =  \min_{\substack{a_1-N^E(p_1) \leq s\leq r }}  \Big(
        cost(a_1-N^E(p_1),s) + f_r(|P|, a_{C_{-1}}, a_{\bar{C}}, r-s)
    \Big)
    \end{multline*}
    for $a_1\geq T_r$ or $a_1=0$.  Otherwise $g(r,a_1,a_{C_{-1}})=\infty$. 

Next, to decide the problem, one iterates through all values $r\in \{ 0,\ldots, |W|\}$, and for each such $r$, for all values of $a_1,a_{C_{-1}},a_{\bar{C}}\in [n+r]$.  If for any such triplet all the objectives hold (that is: $\frac{a_1+a_{C_{-1}}} {a_1+a_{C_{-1}}+a_{\bar{C}}} \geq \varphi, {a_1 \over {a_1+a_{C_{-1}}}} \geq \rho$ and $g(r,a_1,a_{C_{-1}},,a_{\bar{C}})\leq B$), then return ''yes'', otherwise return ''no''.  This completes the algorithm for adding voters.

For the case of deleting voters, first note that the above algorithm also works for negative voter costs.  Hence, we can employ the following simple transformation that reduces an instance of deleting voters to that of adding voters. Given a deletion instance with $V,cost(\cdot), B$, define $\tilde{V}=\emptyset, \tilde{W}=V, \tilde{cost}(v)= -cost(v)$ for all $v$,  and $\tilde{B}=B-cost(V)$.  Now, a solution $\tilde{W}^{+}$ to the addition $\sim$ instance, implies that $V^{-}=V\setminus \tilde{W}^{+}$ is a solution to the original instance. To see this, note that $cost(V^{-})=cost(V)+\tilde{cost}(W^{+})$. Hence, $\tilde{cost}(W^{+})\leq \tilde{B}=B\!-\!cost(V)$ implies $cost(V^{-})\leq B$, as necessary.


\end{proof}


\section{Party Control}\label{sec:party-control}
Party-Control allows the manipulator to change only the set of parties to $\hat{P}$ (as described bellow).
The set of voters remains unchanged, $\hat{V} := V$, and the preferences are $\hat{\succ} := \succ^{\hat{P}}$, where $\succ^{\hat{P}}$ represents the voters’ preferences restricted to $\hat{P}$ (we assume that the original preferences $\succ$ are defined over all parties that \emph{might} run in the election, and that $\hat{\succ}$ coincides with $\succ$ on $\hat{P}$).


Unlike the voter control setting discussed in the previous section, here we focus on a simpler model. Specifically, we assume a uniform cost of 1 for adding or deleting any party, and a cardinality bound $k$ on the number of such actions. This simplification is justified since the problems remain computationally hard even under this simplification.

\subsection{Party Control Types}

\subsubsection{\textbf{Control by Deleting Parties}}
Given an integer $1 \leq k \leq m$, the chair may delete up to $k$ parties from $P$. 

\begin{boxC}{gray}{\textbf{Control by Deleting Parties in Parliamentary Elections}}
\textbf{Input: }\hspace{0.1em} (P, V, $\runningorders$, $k$, $\mathcal{O}$) where:
\begin{tabular}{l l l}
&$P$: Set of $m$ Parties, \quad $V$: Voters, \quad $\runningorders$: Preferences, 
\quad
$k$: Integer  ($1 \leq k \leq m$), \quad\\
&$\mathcal{O}$ : Objective (J or J+F)
\end{tabular}

\tcblower
\textbf{Output: } 
\begin{tabular}{l l }
&Does there exist a set $P^{-} \subseteq \universe$ with $|P^{-}| \leq k$ such that the chair's objective $\mathcal{O}$ is met in the\\
&election $\tilde{E}:=(\hat{P}, V, \bbribe)$ where $\running := \universe \setminus P^{-}$ and $\hat{\succ} := \succ^{\hat{P}}$ (i.e., $\succ$ restricted to $\hat{P}$)?
\end{tabular}
\end{boxC}

\subsubsection{\textbf{Control by Adding Parties}}
We are given a set of spoiler parties, $S$ with $S \cap P=\emptyset$ and $|S|:= \ell$.
Notice that the preference orders are assumed to be over the entire set $P\cup S$.  
Given an integer $1 \leq k \leq \ell$,  the chair may add up to $k$ spoiler parties. 


\begin{boxC}{gray}{\textbf{Control by Adding Parties in Parliamentary Elections}}
\textbf{Input: }\hspace{0.1em} (P, V, $\runningorders$, $k$, $\mathcal{O}$) where:
\begin{tabular}{l l l}
& $P$: Parties, \quad $S$: Set of $\ell$ Spoiler Parties, \quad $V$: Voters, \quad $\runningorders$: Preferences, \\
& $k$: Integer  ($1 \leq k \leq \ell$), \quad $\mathcal{O}$ : Objective (J or J+F)
\end{tabular}

\tcblower
\textbf{Output:} 
\begin{tabular}{l l}
    & Does there exist a set $S^{+} \subseteq S$ with $|S^{+}| \leq k$ such that the chair's objective $\mathcal{O}$ is met in the \\
    & election $\tilde{E}:=(\hat{P}, V, \bbribe)$ where $\running := P\cup S^{+}$ and $\hat{\succ} := \succ^{\hat{P}}$ (i.e., $\succ$ restricted to $\hat{P}$)?
\end{tabular}
\end{boxC}

\subsubsection{\textbf{Party Control Variants}}
We consider separately the cases of adding/deleting coalition parties,
and the cases of adding/deleting opposition parties. 
Thus, we have four separate cases: 
\begin{itemize}
    \item Deleting Coalition Parties (DCP)

    \item Deleting Opposition Parties (DOP)
    
    \item Adding Coalition Parties (ACP).

    \item Adding Opposition Parties (AOP)

\end{itemize}


\subsection{Party Control: Summary of Results}
We consider the parametrized version of the problem, where the parameter is $k$ --- the number of added or deleted parties\footnote{Notice that the problem is trivially FPT in the number of parties.}. The results, summarized in Table \ref{tab:control-results-general}, establish that all relevant cases are hard even in the parametric sense ($W[1]$ or $W[2]$-hard), and, NP-hard. Two cases, \problem{C}{AOP}{J}{=} and \problem{C}{DCP}{J}{=}, are \emph{immune} to manipulation, in the sense that the chair can never advance its objectives by such control.  

\begin{table}[h]
\centering
\small 
\setlength{\tabcolsep}{3pt} 
\begin{tabular}{|c|l|l|l|}
\hline
\multicolumn{2}{|c|}{\textbf{Control Type}} & {\textbf{J Objective}} & \textbf{J+F Objective}\\\hhline{|==|=|=|}
\multirow{2}{*}{DCP} & $\thrs=0$
& Immune (Thm. \ref{thm:-ccdcp-immune}) 
& W[1]-hard (Thm. \ref{thm:dcp-general-CFP}) \\
\cline{2-4}
&  $\thrs > 0$
& W[1]-hard (Thm. \ref{thm:dcp-general-C-t>0}) 
& W[1]-hard (Thm. \ref{thm:dcp-general-C-t>0}$^*$) \\
\hhline{|==|=|=|}
\multicolumn{2}{|c|}{DOP} 
& W[1]-hard (Thm. \ref{thm:dop-general-C}) 
& W[1]-hard (Thm. \ref{thm:dop-general-C}$^*$) \\
\hhline{|==|=|=|}
\multicolumn{2}{|c|}{ACP} 
& W[2]-hard (Thm. \ref{thm:acp-general-C}) 
& W[2]-hard (Thm. \ref{thm:acp-general-C}$^*$) \\
\hhline{|==|=|=|}
\multirow{2}{*}{AOP} & $\thrs=0$
& Immune  (Thm. \ref{thm:-ccaop-immune}) 
& W[2]-hard (Thm. \ref{thm:aop-general-CFP}) \\
\cline{2-4}
 &  $\thrs > 0$
& W[2]-hard (Thm. \ref{thm:aop-general-C-t>0}) 
& W[2]-hard (Thm. \ref{thm:aop-general-C-t>0}$^*$) \\
\hline
\end{tabular}
\caption{Complexity results for party control, parameterized by the number of parties to add or delete. In parentheses --- the corresponding theorem. An asterisk denotes that the result \emph{follows} directly from the theorem.
}
\label{tab:control-results-general}
\end{table}

\subsection{Deleting Coalition Parties (DCP)}
When $\thrs=0$, \Cref{thm:-ccdcp-immune} proves that deleting coalition parties cannot advance the J objective, as such deletion can only decrease the coalition's share. 
In contrast, when a favored party is also taken under consideration ($\rho >0$), \Cref{thm:dcp-general-CFP} shows that the objectives might be achieved, but it is W[1]-hard.




When $\thrs>0$, \Cref{thm:dcp-general-C-t>0} proves that the problem is W[1]-hard even when $\rho =0$.

\begin{theorem}\label{thm:-ccdcp-immune}
      \problem{C}{DCP}{J}{=} is not possible.
\end{theorem} 
\begin{proof}
    Each voter gives exactly one point to the party ranked first.  
    Since all deleted parties are from the coalition, we distinguish between two cases:
    \begin{itemize}
        \item  For a voter whose first-ranked party is not from the coalition, deleting a coalition party, does not change the outcome, and the total points received by opposition parties remain the same.
        \item For a voter whose first-ranked party is from the coalition, deleting a coalition party that is in the first position may reduce the points received by the coalition, if the next party is from the opposition.  
    \end{itemize}
    
    This problem concerns only the percentage of points received by the coalition as a whole, rather than by any specific favored party within it.
    As a result, deleting opposition parties can only harm the coalition's overall standing.
\end{proof}

\begin{theorem} \label{thm:dcp-general-CFP}
    \problem{C}{DCP}{J+F}{=} parametrized by the number of deleted parties is W[1]-hard.
\end{theorem}

\begin{proof}
    The proof is by reduction from The $k$-Clique Problem: Given an undirected graph $G=(W, E)$, and a parameter $k$, it determines whether there is a clique of size exactly $k$. 

    We show that given an instance to The $k$-Clique Problem, we can construct an instance of the \problem{C}{DCP}{J+F}{=} problem with the same parameter, $k$, such that a $k$-clique exists if and only if achieving the J+F objective by deleting $k$ coalition parties is possible. 

    Let $\universe:=\{p_1, o_1,~ s_1,\ldots, s_{|W|}\}$, where each $s_i$ is corresponding to a vertex in $G$; $C:=\{p_1, s_1,\ldots, s_{|W|}\}$, and the preferred party $p_1$; $\varphi := \frac{1}{2}$, and $\rho := \frac{|E|+\binom{k}{2}}{2|E|}$.

    There are $4|E|$ voters, $4$ voters corresponding to each edge $(x,y) \in E$, whose preferences are defined as follows:  
    \begin{itemize}
        \item $1$ voter of Type $A$:~~~ $s_x \succeq s_y \succ p_1 \succ o_1 \succ \ldots$
        
        \item $1$ voter of Type $B$:~~~ $p_1 \succ o_1 \succ \ldots $
        
        \item $2$ voter of Type $C$:~~~ $o_1 \succ \ldots$
    \end{itemize}
    
    ($\Rightarrow$) Assume there exists a $k$-clique in the graph. We prove that the J+F objective can be achieved by removing the $k$ coalition parties corresponding to the vertices in the clique.
    
    By definition, there are $\binom{k}{2}$ edges whose both endpoints lie within the clique. 
    Thus, after removing these parties: $p_1$ gets $|E|+\binom{k}{2}$ votes, the set of parties corresponding to $W$ gets $|E|-\binom{k}{2}$ voters; and $o_1$ gets  $2|E|$ votes.
    Hence, the coalition achieve $\frac{1}{2}$, and $p_1$ achieve $\frac{|E|+\binom{k}{2}}{2|E|}$ of the coalition's votes as required.

    ($\Leftarrow$) Assume that the J+F objective can be achieved by removing $k$ coalition parties. We prove that the the corresponding vertices is a $k$-clique.

    In the original election, $o_1$ gets $2|E|$ votes. As all voters whose top choice is not $o_1$ prefer $p_1$ over $o_1$, removing $C_{-1}$ parties cannot affect $o_1$'s vote count.
    This implies that the coalition's vote count remains $2|E|$ for any removal of coalition parties different than $p_1$.
    Since we know that the favorite-party objective (F) is satisfied for $\rho = \frac{|E|+\binom{k}{2}}{2|E|}$, $p_1$, must get at least $|E|+\binom{k}{2}$ votes.
    In the original election, $p_1$ receives $|E|$ votes, so the party deletions contribute at least $\binom{k}{2}$ additional votes to $p_1$.
    
    ~Each such new vote for $p_1$ can only come from a voter of Type A, whose  top two parties correspond to vertices $x$ and $y$.
    To gain at least $\binom{k}{2}$ new votes in this manner, the deleted parties must correspond to a set of $k$ vertices that induce at least $\binom{k}{2}$ edges. 
    This is only possible if these vertices form a clique of size $k$.
\end{proof}

\begin{theorem}\label{thm:dcp-general-C-t>0}
    \problem{C}{DCP}{J}{\geq} parametrized by the
number of  deleted parties is W[1]-hard.
\end{theorem}
\begin{proof}
    The proof is by reduction from the $k$-clique problem: Given an undirected graph $G=(W, E)$, and a parameter $k$, it determines whether there is a clique of size exactly $k$. 
 
    We show that given an instance to the $k$-clique problem, we can construct an instance of the \problem{C}{DCP}{J}{\geq} problem with the same parameter, $k$, such that a $k$-clique exists if and only if achieving the J objective by deleting $k$ coalition parties is possible.
    Let $\universe:=\{p_1, o_1, ~s_1,\ldots, s_n\}$, where each $s_i$ is corresponding to a vertex in $G$;
     $C=\{p_1, s_1,\ldots, s_n\}$;
     $\varphi = \frac{1}{2}$, and the election threshold $\thrs=\frac{|E|+\binom{k}{2}}{3|E|+\binom{k}{2}}$.
    %
    There are $3|E| + \binom{k}{2}$ voters, whose preferences are defined as follows:
    \begin{itemize}
         \item $|E|$ voters of Type A: each voter is corresponding to an edge $(x,y) \in E$  in the graph. Their preferences are $s_x \succ s_y \succ p_1 \succ o_1 \succ P\ldots$
         
         \item $|E|$ voters of Type B: $p_1 \succ o_1 \succ \ldots$
         
         \item $|E|$ voters of Type C: $o_1 \succ \ldots$
         
         \item $\binom{k}{2}$ voters of Type D: $o_1 \succ \ldots$
        \end{itemize}


    
    
    ($\Rightarrow$) Assume there exists a $k$-clique in the graph.  We prove that the J objective can be achieved by removing the $k$  parties corresponding to the vertices in the clique.
     By definition, there are $\binom{k}{2}$ edges whose both endpoints lie within the clique. Thus, after removing these parties, the first party will be $p_1$, so $p_1$ will get $|E|+\binom{k}{2}$ votes.
    In the other $|E|-\binom{k}{2}$ voters $v_{1, i}$, the first party is not $p_1$, but it is an inactive vote of the coalition.
    In addition, $o_1$, which is not part of the coalition, gets  $|E|+\binom{k}{2}$ votes, and it cannot be changed by deletion of parties from the coalition.
    Hence, there are $2(|E|+\binom{k}{2})$ active votes and the coalition achieve $\varphi = \frac{1}{2}$. 
     
    ($\Leftarrow$) Assume that the J objective can be achieved by removing $k$ parties.  
    In the original election, $o_1$, which is not part of the coalition, gets  $|E|+\binom{k}{2}$ votes, and it cannot be changed by deleting parties from the coalition.
    The parties $s_1,\ldots, s_{|W|}$ are part of the coalition, but none of them passed the election threshold, and it cannot be changed by deleting parties from the coalition.
    In order to satisfy the coalition objective, $p_1$, must get at least $|E|+\binom{k}{2}$ votes, that is, achieve at least $\binom{k}{2}$ new votes by the control action. 
    If a deletion results in a vote for $p_1$, it implies that both vertices corresponding to the endpoints of the associated edge are among the deleted parties.  
    This means the deletion of $k$ parties successfully includes the vertices representing both endpoints of the edges for $\binom{k}{2}$ edges.  
    Consequently, all edges between these vertices must exist in the graph $G$, indicating that these vertices form a clique.     
\end{proof}


\subsection{Deleting Opposition Parties (DOP)}

\Cref{thm:dop-general-C} proves that the problem is W[1]-hard even when $\rho =0$ and the election threshold $\thrs=0$.

\begin{theorem} \label{thm:dop-general-C}
     \problem{C}{DOP}{J}{=} parametrized by the number of  deleted parties is W[1]-hard.
\end{theorem}
\begin{proof}
    The proof is by reduction from the $k$-clique problem: Given an undirected graph $G=(W, D)$, and a parameter $k$, it determines whether there is a clique of size exactly $k$. 
 
    We show that given an instance to the $k$-clique problem, we can construct an instance of the \problem{C}{DOP}{J}{=} problem with the same parameter, $k$, such that a $k$-clique exists if and only if achieving the J objective by deleting $k$ opposition parties is possible.
    

    Let $P=\{p_1,s_1,\ldots, s_{|W|}\}$, each $s_i$ is corresponding to a vertex in $G$.
    Let the coalition to be $C=\{p_1\}$. 
    Let the parameter of the number of deleting parties be $k$.
    Let $\varphi = \frac{\binom{k}{2}}{|E|}$ (note that, if $\varphi>1$, then there is no clique of size $k$ in the graph since there are no $\binom{k}{2}$ edges at all).
    %
    Let $V = \{v_1,\ldots v_{|E|}\}$  be the set of voters, each voter corresponds to one edge in $G$. 
    The  preference order of $v_i$, that corresponds to the edge $(x,y)$ is defined to be $s_x \succ s_y \succ p_1 \succ P\setminus \{s_x,s_y,p_1\}$.

    ($\Rightarrow$) Assume there exists a k-clique in the graph, then delete the corresponding $k$ parties.
    Each edge between these vertices is in the graph, hence, there are exactly $\binom{k}{2}$ such edges. Thus, in the preference order of all voters that correspond to these edges, after the deletion, the first party will be $p_1$, and the coalition achieves $\frac{\binom{k}{2}}{|E|}$ as desired.

    ($\Leftarrow$) Suppose a successful control action exists.  
    In the original election, $p_1$ does not receive any votes, and there are $|E|$ voters participating.  
    Thus, the control action must secure at least $\binom{k}{2}$ votes by deleting $k$ parties.  
    If a deletion results in a vote for $p_1$, it implies that both vertices corresponding to the endpoints of the associated edge are among the deleted parties.  
    This means the deletion of $k$ parties successfully includes the vertices representing both endpoints of the edges for $\binom{k}{2}$ edges.  
    Consequently, all edges between these vertices must exist in the graph $G$, indicating that these vertices form a clique. 
\end{proof}


\subsection{Adding Coalition Parties (ACP)}
    Party-Control for a Coalition by Adding Coalition Parties is W[2]-hard, even when $\thrs=0$.
    Being the simplest variant, this suggests that the remaining variants are at least as hard.

\begin{theorem}\label{thm:acp-general-C}
\problem{C}{ACP}{J}{=} parametrized by the number of added parties is W[2]-hard.
\end{theorem}




\begin{proof}
    The proof is by reduction from The Dominating Set Problem: Given an undirected graph $G=(W, E)$, and a parameter $k$, it determines whether there is a dominating set of size at most $k$. 

    We show that given an instance to The Dominating Set Problem, we can construct an instance of the \problem{C}{ACP}{J}{=} problem with the same parameter, $k$, such that a dominating set of size at most $k$ exists if and only if achieving the J objective by adding $k$ coalition parties is possible. 
    
    Let the set of spoiler parties be $S:=\{s_1,\ldots, s_{|W|}\}$, where each spoiler party $s_i$ corresponds to the vertex $w_i\in W$ of $G$.
    Let $\universe := \{p_1,o_2\}$.
    Let $N[s_i]$ be the set of parties corresponding to $w_i$'s neighbors --- that is, 
    $ N[s_i] := \{s_j \in S \mid (w_i,w_j)\in D\}$.
    There are $2|W|$ voters, $2$ voters corresponding to each vertex $w_i \in W$, whose preferences are defined as follows:  
    \begin{itemize}
        \item $1$ voter of Type $A$:~~~ its most favorite parties are those in $\{s_i\} \cup N[s_i]$ (the order between them does not matter); then $o_2$, then $p_1$, and then the rest of the parties. 
        
        \item $1$ voter of Type $B$:~~~ $o_2 \succ p_1 \succ \ldots$ 
    \end{itemize}

    Let the coalition be $C := \{p_1, ~ s_1,\ldots, s_{|W|}\}$ and the target-fraction be $\varphi := \frac{1}{2}$.

    
    ($\Rightarrow$)
    Assume there exists a dominating set of size at most $k$.
    In this case, the control action that adds the corresponding spoiler parties ensures that the coalition will achieve $50\%$ of the votes.
    This is because, for voters of type A, the first party in each preference order of voter $v_i$ is one of the spoiler parties that belong to the coalition and is part of the dominating set.
    For voters of type B, the first party in each preference order of voter $v_i$ is $o_2$, which is not part of the coalition, and it cannot be changed by adding spoiler parties.
    As a result, the coalition achieves $50\%$ of the votes.

    ($\Leftarrow$) Assume that there is a successful control action, $S'\subseteq S$, that added at most $k$ spoiler parties and achieves $50\%$ of the votes for the coalition. 
    For voters of type B, the first party in each preference order of voter $v_i$ is $o_2$, which is not part of the coalition, and it cannot be changed by adding parties.
    Hence, it must be that for all voters of type A, the first party in each preference order of voter $v_i$ is part of the coalition. 
    That is, for each voter one of $N[s_i]$ must be part of $S'$, and $|S'|\leq k$, hence the set that contains the corresponding vertices is a dominating set of the given graph.
\end{proof}


\subsection{Adding Opposition Parties (AOP)}

When $\thrs=0$, \Cref{thm:-ccaop-immune} proves that adding opposition parties cannot advance the J objective, as such addition can only increase the opposition's share. 
However, when  $\thrs>0$, or for the C+F objective, the problems are W[2]-hard (\Cref{thm:aop-general-C-t>0} and \Cref{thm:aop-general-CFP}, respectively). 


\begin{theorem}\label{thm:-ccaop-immune}
    \problem{C}{AOP}{J}{=} is immune.
\end{theorem} 

\begin{proof}
    Each voter gives exactly one point to the party ranked first; hence, adding parties to other positions makes no difference.
    We distinguish between two possible cases:
    \begin{itemize}
        \item For a voter whose top choice is outside the coalition, placing a spoiler opposition party first does not affect the outcome, as the opposition's total points remain unchanged.

        \item For a voter whose top choice is in the coalition, placing a spoiler opposition party first reduces the coalition's points. 
    \end{itemize}
    
    This problem concerns only the percentage of points received by the coalition as a whole, rather than by any specific favored party within it.
    As a result, adding spoiler opposition parties can only harm the coalition's overall standing.
\end{proof}

\begin{theorem}\label{thm:aop-general-C-t>0}
\problem{C}{AOP}{J}{\geq} parametrized by the number of added parties is W[2]-hard.
\end{theorem}

The proof adopts a similar approach as that of \cite{liu2009parameterized}. 

\begin{proof}
    The proof is by reduction from The Dominating Set Problem: Given an undirected graph $G=(W, E)$, and a parameter $k$, it determines whether there is a dominating set of size at most $k$. 

    We show that given an instance to The Dominating Set Problem, we can construct an instance of the \problem{C}{AOP}{J}{\geq} problem with the same parameter, $k$, such that a dominating set of size at most $k$ exists if and only if achieving the J objective by adding $k$ coalition parties is possible. 

    
    Let the set of spoiler parties $S=\{s_1,\ldots, s_{|W|}\}$ contain a spoiler party for each vertex in $G$.
    Let $\universe= \{p_1,o_2, o_3\} \cup S$ be the set of parties.  
    Let the coalition be $C=\{p_1\}$.
    Let $\varphi = 50\%$. Let the election threshold $\thrs=40\%$.
    Let $N[s_i]$ be the set of parties corresponding to $w_i$'s neighbors --- that is, 
    $ N[s_i] := \{s_j \in S \mid (w_i,w_j)\in D\}$.
    There are $5|W|$ voters, $5$ voters corresponding to each vertex $w_i \in W$, whose preferences are defined as follows:  
    \begin{itemize}
         \item $2$ voter of Type $A$:~~~ $p_1 \succ o_2 \succ \ldots$

         \item $2$ voter of Type $B$:~~~ $o_2 \succ p_1 \succ \ldots$
    
        \item $1$ voter of Type $C$:~~~ its most favorite parties are those in $\{s_i\} \cup N[s_i]$ and it is indifference between them; then $o_2$, then $p_1$, and then the rest of the parties. 
    \end{itemize}
    
     Notice that before any control action, the coalition achieves $40\%$
 
   ($\Rightarrow$)
    Assume there exists a dominating set of size at most $k$.
    In this case, the control action that adds the corresponding spoiler parties ensures that the coalition will achieve $50\%$ of the votes.

    This is because, there are $4|W|$ voters, those of types A and B, whose favorite party  is not one of the spoiler parties and it cannot be changed by control by adding parties.
    After the control action, for all $|W|$ voters of type C, their favorite party is a spoiler one that does not pass the election threshold. 
    As a result, only $4|W|$ votes are active after the control action, and the coalition achieves $50\%$ of the votes, as required.

    ($\Leftarrow$) Assume that there is a successful control action, $S'\subseteq S$, that added at most $k$ spoiler parties and achieves a successful control action.
    It is not possible to achieve more votes to $p_1$, so the only way to meet the requirement of $\varphi$ is by reducing the size of the opposition.
    It is not possible to change the favorite party of all $4 |W|$ voters of Type A and B by control by adding parties.

    Hence, it must be that the control action affect the favorite party of voter of Type C which is an active vote to the opposition before the control action become an inactive vote to the opposition after the control action. 
    That is, for each voter, $v_i$, one of $\{s_i\} \cup N[s_i]$ must be part of $S'$, and $|S'|\leq k$, hence the set that contains the corresponding vertices is a dominating set of the given graph.
    
\end{proof}

\begin{theorem} \label{thm:aop-general-CFP}
\problem{C}{AOP}{J+F}{=} parametrized by the number of added parties is W[2]-hard.
\end{theorem}

\begin{proof}
    The proof is by reduction from the Dominating Set Problem: Given an undirected graph $G=(W, E)$, and a parameter $k$, it determines whether there is a dominating set of size at most $k$. 

    We show that given an instance to The Dominating Set Problem, we can construct an instance of the \problem{C}{AOP}{J+F}{=} problem with the same parameter, $k$, such that a dominating set of size at most $k$ exists if and only if achieving the J+F objective by adding $k$ coalition parties is possible. 

    
    Let the set of spoiler parties $S=\{s_1,\ldots, s_{|W|}\}$ contain a spoiler party for each vertex in $G$.
    Let $\universe= \{p_1,c_2\}$, the coalition be $C=\{p_1,c_2\}$, and the favored party be $p_1$.
    Let the parameter of the number of added parties be $k$.
    Let $\varphi = 50\%$, and let $\rho = 50\%$.

    Let $N[s_i]$ be the set of parties corresponding to $w_i$'s neighbors --- that is, 
    $ N[s_i] := \{s_j \in S \mid (w_i,w_j)\in D\}$.

    There are $4|W|$ voters, $4$ voters corresponding to each vertex $w_i \in W$, whose preferences are defined as follows:  
    \begin{itemize}
        \item $1$ voter of Type $A$:~~~ $p_1 \succ c_2 \succ\ldots$
        
        \item $1$ voter of Type $B$:~~~ $c_2 \succ p_2 \succ \ldots$
    
        \item $2$ voter of Type $C$:~~~ its most favorite parties are those in $\{s_i\} \cup N[s_i]$ and it is indifference between them; then $c_2$, then $p_1$, and then the rest of the parties. 
    \end{itemize}
    
 
    Notice that before any control action, the coalition achieves $100\%$ of the votes, but, $p_1$ achieves only $0.25\%$ of the coalition.

    ($\Rightarrow$)
    Assume there exists a dominating set of size at most $k$.
    In this case, the control action that adds the corresponding spoiler parties ensures that the coalition will achieve only $50\%$ of the votes.
    This is because, for all $2|W|$ voters of Types A and B, their favorite party is not one of the spoiler parties and it cannot be changed by control by adding parties.
    After the control action, for the $2|W|$ voters of Type C, their favorite party is a spoiler one that is not part of the coalition. 
    As a result, the coalition achieves $50\%$ of the votes, and the percentage of $p_1$ among the coalition becomes $0.5\%$ as required.

    ($\Leftarrow$) Assume that there is a successful control action, $S'\subseteq S$, that added at most $k$ spoiler parties and achieves a successful control action.
    It is not possible to achieve more votes to $p_1$, so the only way to meet the requirement of $\rho$ is by reducing the size of the coalition, without reducing the number of votes to $p_1$.
    For the $2|W|$ voters of Types A and B, their favorite party cannot be changed by control by adding parties.

    Hence, it must be that for the $2|W|$ voters of Type C, their favorite party is not part of the coalition after the control action. 
    That is, for each voter one of $\{s_i\} \cup N[s_i]$ must be part of $S'$, and $|S'|\leq k$, hence the set that contains the corresponding vertices is a dominating set of the given graph.
\end{proof}



\section{Future Work}\label{sec:conclusion}

We introduced and studied bribery and control in parliamentary elections,  when the goal is to promote an entire coalition of parties. 
This work lays the foundation for a wide range of open questions.

 


\paragraph{Manipulation Types}
Broadening the scope of strategic manipulations in parliamentary elections is especially compelling—particularly the study of control via party splits and mergers, a practice frequently observed in reality.
Additionally, changing the electoral threshold can have dramatic effects and is commonly used in practice.  Accordingly,
manipulation of the electoral threshold is a natural and interesting research direction.
Another direction is destructive manipulations, where the manipulator aims to prevent a rival-coalition from succeeding. 

\paragraph{Coalition Dynamics}
In this paper, we considered a single party and a single coalition of parties that the manipulator seeks to promote.  In practice, political dynamics are richer and more complex.  For example, the manipulator may be interested in promoting one of several possible coalitions --- e.g.~ any center/left/right coalition.  Alternatively, the manipulator may seek to guarantee that its preferred party is part of \emph{some} winning coalition. Likewise, on the destructive end, the manipulator may aim that a specific party, or parties, are \emph{not} part of any winning coalition.  
It would also be interesting to study scenarios where the manipulator not only has one favored party but specifies a desired balance of power among the coalition parties.

\paragraph{Multiple Manipulators}
In this paper, we assumed a single manipulator and the question was only a computational one.  In practice, there may be several manipulators, each with its own goals and objectives.  In this case, \emph{strategic} considerations come into play.  The study of such dynamics, and equilibrium thereof is an interesting future direction.




\section{Acknowledgments}
This research is partly supported by the Israel Science Foundation grants 2544/24, 3007/24 and 2697/22.


\clearpage
\bibliographystyle{elsarticle-num-names}
\bibliography{main}

\clearpage
\appendix

\setcounter{secnumdepth}{2} 
\renewcommand\thesection{\Alph{section}}

\clearpage
\section{Known NP-Hard Problems Used}\label{apx:np-probs}
\begin{definition}[3-4-Exact-Cover problem]
    \textbf{Given:} A set $Z$ of elements.
    A collection $\mathcal{D}$ of $4$-subsets of $Z$ (i.e., for any $D \in \mathcal{D}$: $D\subseteq Z$ and $|D|=4$); such that each element $z \in Z$ belongs to exactly $3$ subsets in $\mathcal{D}$.
    \textbf{Question:} Does there exists a
    a sub-collection $\mathcal{D}'\subseteq \mathcal{D}$ such that any element in $Z$ appears exactly in one subset in $\mathcal{D}'$.
\end{definition}
The 3-4-Exact-Cover Problem is NP-Hard
\cite{brelsford2008approximability}.

\begin{definition}[Dominating Set Problem]
    \textbf{Given:} A graph $G = (W,E)$, and a parameter $k$.
    \textbf{Question:} Does there exists a dominating set $W'\subseteq W$ of size at most $k$ --- i.e., for any vertex $w\in W$, either $w$ is in the cover, $w \in W'$; or that one of its neighbors is in the cover, $\exists w'\in W'\ s.t\ (w,w')\in E$.
\end{definition}
The Dominating Set Problem is W[2]-hard \cite{downey1999structure}.

\begin{definition}[k-Clique problem]
    \textbf{Given:} A graph $G = (W,E)$, and a parameter $k$.
    \textbf{Question:} Does there exists a clique of size $k$ in the graph. That is a set, $W'$, of $k$ vertices such that $\forall w,w'\in W' (w,w')\in E$.
\end{definition}
The k-clique problem is a W[1]-hard problem \cite{Fixed1995Downey}.

\end{document}

%% file: includes-and-macros.tex
\usepackage{bm}
\usepackage{latexsym}
\usepackage{amsmath}
\usepackage{amsthm}
\usepackage{booktabs}
\usepackage{enumitem}
\usepackage{graphicx}
\usepackage{color}
\usepackage{tabularx}
\usepackage{cleveref}
\usepackage{algorithm}
\usepackage[noend]{algorithmic}
\usepackage{array}
\usepackage{hhline}

\newtheorem{theorem}{Theorem}

\newtheorem{definition}{Definition}
\newcounter{examplecounter}

\newtheorem*{rtheorem}{\theremindertheorem}

\newcommand{\theremindertheorem}{}

\usepackage{mathtools}
\usepackage{mathrsfs}
\usepackage[table, dvipsnames]{xcolor}

\newcommand{\oi}{\succ}
\newcommand{\oit}{\succ}
\newcommand{\sbribe}{\mathrel{\hat{\oit}}}
\newcommand{\universeOi}{\oit}
\newcommand{\order}{\bm{\succ}}
\newcommand{\ordert}{\bm{\succ}}
\newcommand{\universeorders}{{\bm{\succ}}}
\newcommand{\bbribe}{\hat{\ordert}}
\newcommand{\runningorders}{{\ordert}}

\newcommand{\universe}{P}
\newcommand{\running}{\hat{P}}

\newcommand{\universeVoters}{V}
\newcommand{\runningVoters}{V}

\newcommand{\spoilerVoters}{W}
\newcommand{\controlV}{\hat{V}}
\newcommand{\influencedE}{\hat{E}}
\newcommand{\pos}[3]{\text{Pos}^{#1}(#2,#3)}
\newcommand{\topT}[2]{\mbox{\it Top}^{#1}(#2)}
\newcommand{\Npoint}[2]{N^{#1}(#2)}
\newcommand{\seats}[2]{\mathrm{frac}_{#1}^{#2}}
\newcommand{\costIbribery}[3]{\textit{Cost}_{#1}(#2,#3)}
\newcommand{\costIbriberyshort}[2]{\textit{Cost}_{#1}(#2)}
\newcommand{\costbribery}[2]{\textit{Cost}(#1,#2)}
\newcommand{\costVoterControl}[1]{\textit{Cost}(#1)}
\newcommand{\costOneB}[3]{\textit{Cost}^1_{#1}(#2,#3)}
\newcommand{\costDolarB}[3]{\textit{Cost}^{\$}_{#1}(#2,#3)}
\newcommand{\costDolarBSet}[1]{\textit{Cost}^{\$}(#1)}
\newcommand{\costShiftB}[3]{\textit{Cost}^{shift}_{#1}(#2,#3)}
\newcommand{\costSwapB}[3]{\textit{Cost}^{swap}_{#1}(#2,#3)}

\newcommand{\thrs}{\tau}
\newcommand{\problemdef}[4]{$\langle$#1$|$#2$|$#3$|$#4$\rangle$}
\newcommand{\problem}[4]{$\langle$#1$|$#2$|$#3$|$$\thrs\!#4\!0\rangle$}

\newcommand{\prefferedParty}{p_1}

\definecolor{mygray}{gray}{0.95}
\definecolor{mygray2}{gray}{0.85}

\usepackage[many]{tcolorbox}

\newtcolorbox{boxB}{
    boxrule = 1.5pt,
    rounded corners,
    arc = 5pt   
}

\newtcolorbox{mybox}[2][]{%
attach boxed title to top center
               = {yshift=-8pt},
  colback      = black!5!white,
  colframe     = black!75!black,
  fonttitle    = \bfseries,
  colbacktitle=white!80!white,
  coltitle = gray!10!black,
  title        = #2,#1,
  enhanced,
}

\newtcolorbox[auto counter, number within=section]{myboxA}[2][]{enhanced, 
  attach boxed title to top center={yshift=-3.5mm,yshifttext=-1mm},
  colback=white,
  colframe=black,
  colbacktitle=white,
  fonttitle=\bfseries,
  coltitle=black,
  boxed title style={colframe=white, coltext=black},
  title=#2, #1}

\newtcolorbox{boxC}[3][left=2pt, right=2pt]
{
  colframe = #2!25,
  colback  = #2!5,
  coltitle = #2!20!black,  
  title    = {#3},
  #1,
}

 \usepackage{multirow}

 \usepackage[bibliography=common]{apxproof}
 
\newtheoremrep{theorem}{Theorem}[section]
\newtheoremrep{proposition}[theorem]{Proposition}
\newtheoremrep{lemma}[theorem]{Lemma}
\newtheoremrep{claim}[theorem]{Claim}
\newtheoremrep{observation}[theorem]{Observation}
\newtheoremrep{open}[theorem]{Open Question}
\renewcommand{\appendixprelim}{\clearpage\twocolumn}

\renewcommand{\appendixprelim}[1]{%
}

\newcommand{\red}[1]{\textcolor{red}{#1}}

\usepackage{diagbox}

\usepackage{caption}

%% file: example-table.tex
\begin{table*}[h]
\centering
\begin{tabular}{|c|c|c|c|c|c|c|c|c|c|c|c|}
\hline \parbox[c][0.7cm][c]{2.5cm}{}
& & $p_1$ & $c_1$ & $c_2$ & $o_1$ & $o_2$ & $\seats{\thrs}{\hat{E}}(p_1)$ & $\seats{\thrs}{\hat{E}}(C)$ & 
$ \displaystyle \frac{\seats{\thrs}{\hat{E}}(p_1)}{ \seats{\thrs}{\hat{E}}(C)}$
& \ref{eq:target-coalition-obj} & \ref{eq:target-favored-party-obj}\\
\hhline{|=|=|=|=|=|=|=|=|=|=|=|=|}
\multirow{2}{*}{no manipulation} & votes & 20 & 10 & 0 & 25 & 20 & \multirow{2}{*}{31\%} & \multirow{2}{*}{31\%} & \multirow{2}{*}{100\%} & \multirow{2}{*}{ \textbf{\texttimes} 
} & \multirow{2}{*}{ $\checkmark$ } \\
\cline{2-7}
                    & active votes & 20 & 0 & 0 & 25 & 20 &&&                     &                     &                     \\
\hhline{|=|=|=|=|=|=|=|=|=|=|=|=|}
\multirow{2}{*}{removing $o_1$} & votes & 25 & 30 & 0 & 0 & 20 & \multirow{2}{*}{33\%} & \multirow{2}{*}{73\%} & \multirow{2}{*}{45\%} & \multirow{2}{*}{ $\checkmark$ 
} & \multirow{2}{*}{ \textbf{\texttimes} } \\
\cline{2-7}
                    & active votes & 25 & 30 & 0 & 0 & 20 &                     &&&                     &                     \\
\hhline{|=|=|=|=|=|=|=|=|=|=|=|=|}
\multirow{2}{*}{removing $o_2$} & votes & 20 & 10 & 20 & 25 & 0 & \multirow{2}{*}{31\%} & \multirow{2}{*}{62\%} & \multirow{2}{*}{50\%} & \multirow{2}{*}{ $\checkmark$ 
} & \multirow{2}{*}{ $\checkmark$}\\
\cline{2-7}
                    & active votes & 20 & 0 & 20 & 25 & 0 &                &&     &                     &                     \\
\hhline{|=|=|=|=|=|=|=|=|=|=|=|=|}
\multirow{2}{*}{removing $c_1$} & votes & 20 & 0 & 0 & 35 & 20 & \multirow{2}{*}{27\%} & \multirow{2}{*}{27\%} & \multirow{2}{*}{100\%} & \multirow{2}{*}{ \textbf{\texttimes} 
} & \multirow{2}{*}{ $\checkmark$ }\\
\cline{2-7}
                    & active votes & 20 & 0 & 0 & 35 & 20 &       &&              &                     &                     \\
\hhline{|=|=|=|=|=|=|=|=|=|=|=|=|}
\multirow{2}{*}{removing $c_2$} & votes & 20 & 10 & 0 & 25 & 20 & \multirow{2}{*}{31\%} & \multirow{2}{*}{31\%} & \multirow{2}{*}{100\%} & \multirow{2}{*}{ \textbf{\texttimes} 
} & \multirow{2}{*}{ $\checkmark$ } \\
\cline{2-7}
                    & active votes & 20 & 0 & 0 & 25 & 20 &       &&              &                     &                     \\
\hline
\end{tabular}
\caption{Example: Effects of removing parties\label{tab:example}. The election threshold is $\thrs=15\%$, and both the coalition target-fraction and the target-favored-party-ratio are $\varphi=\rho=50\%$. } 
\end{table*}

%% file: main.bbl
\begin{thebibliography}{45}
\expandafter\ifx\csname natexlab\endcsname\relax\def\natexlab#1{#1}\fi
\providecommand{\url}[1]{\texttt{#1}}
\providecommand{\href}[2]{#2}
\providecommand{\path}[1]{#1}
\providecommand{\DOIprefix}{doi:}
\providecommand{\ArXivprefix}{arXiv:}
\providecommand{\URLprefix}{URL: }
\providecommand{\Pubmedprefix}{pmid:}
\providecommand{\doi}[1]{\href{http://dx.doi.org/#1}{\path{#1}}}
\providecommand{\Pubmed}[1]{\href{pmid:#1}{\path{#1}}}
\providecommand{\bibinfo}[2]{#2}
\ifx\xfnm\relax \def\xfnm[#1]{\unskip,\space#1}\fi
\bibitem[{Betzler and Uhlmann(2009)}]{betzler2009parameterized}
\bibinfo{author}{N.~Betzler}, \bibinfo{author}{J.~Uhlmann},
\newblock \bibinfo{title}{Parameterized complexity of candidate control in elections and related digraph problems},
\newblock \bibinfo{journal}{Theoretical Computer Science} \bibinfo{volume}{410} (\bibinfo{year}{2009}) \bibinfo{pages}{5425--5442}.
\bibitem[{Elkind et~al.(2009)Elkind, Faliszewski, and Slinko}]{elkind2009swap}
\bibinfo{author}{E.~Elkind}, \bibinfo{author}{P.~Faliszewski}, \bibinfo{author}{A.~Slinko},
\newblock \bibinfo{title}{Swap bribery},
\newblock in: \bibinfo{booktitle}{Algorithmic Game Theory: Second International Symposium, SAGT 2009, Paphos, Cyprus, October 18-20, 2009. Proceedings 2}, \bibinfo{organization}{Springer}, \bibinfo{year}{2009}, pp. \bibinfo{pages}{299--310}.
\bibitem[{Elkind et~al.(2020)Elkind, Faliszewski, Gupta, and Roy}]{elkind2020algorithms}
\bibinfo{author}{E.~Elkind}, \bibinfo{author}{P.~Faliszewski}, \bibinfo{author}{S.~Gupta}, \bibinfo{author}{S.~Roy},
\newblock \bibinfo{title}{Algorithms for swap and shift bribery in structured elections},
\newblock in: \bibinfo{booktitle}{Proceedings of the 19th International Conference on Autonomous Agents and MultiAgent Systems}, AAMAS '20, \bibinfo{publisher}{International Foundation for Autonomous Agents and Multiagent Systems}, \bibinfo{address}{Richland, SC}, \bibinfo{year}{2020}, p. \bibinfo{pages}{366–374}.
\bibitem[{Faliszewski and Rothe(2016)}]{faliszewski2016control}
\bibinfo{author}{P.~Faliszewski}, \bibinfo{author}{J.~Rothe},
\newblock \bibinfo{title}{Control and bribery in voting},
\newblock in: \bibinfo{editor}{F.~Brandt}, \bibinfo{editor}{V.~Conitzer}, \bibinfo{editor}{U.~Endriss}, \bibinfo{editor}{J.~Lang}, \bibinfo{editor}{A.~D. Procaccia} (Eds.), \bibinfo{booktitle}{Handbook of Computational Social Choice}, \bibinfo{publisher}{Cambridge University Press}, \bibinfo{year}{2016}, pp. \bibinfo{pages}{146--168}. \URLprefix \url{https://doi.org/10.1017/CBO9781107446984.008}. \DOIprefix\doi{10.1017/CBO9781107446984.008}.
\bibitem[{Faliszewski et~al.(2006)Faliszewski, Hemaspaandra, and Hemaspaandra}]{faliszewski2006complexity}
\bibinfo{author}{P.~Faliszewski}, \bibinfo{author}{E.~Hemaspaandra}, \bibinfo{author}{L.~A. Hemaspaandra},
\newblock \bibinfo{title}{The complexity of bribery in elections},
\newblock in: \bibinfo{booktitle}{Proceedings of the 21st National Conference on Artificial Intelligence - Volume 1}, AAAI'06, \bibinfo{publisher}{AAAI Press}, \bibinfo{year}{2006}, p. \bibinfo{pages}{641–646}.
\bibitem[{Faliszewski et~al.(2009)Faliszewski, Hemaspaandra, and Hemaspaandra}]{faliszewski2009hard}
\bibinfo{author}{P.~Faliszewski}, \bibinfo{author}{E.~Hemaspaandra}, \bibinfo{author}{L.~A. Hemaspaandra},
\newblock \bibinfo{title}{How hard is bribery in elections?},
\newblock \bibinfo{journal}{Journal of artificial intelligence research} \bibinfo{volume}{35} (\bibinfo{year}{2009}) \bibinfo{pages}{485--532}.
\bibitem[{Keller et~al.(2018)Keller, Hassidim, and Hazon}]{keller2018approximating}
\bibinfo{author}{O.~Keller}, \bibinfo{author}{A.~Hassidim}, \bibinfo{author}{N.~Hazon},
\newblock \bibinfo{title}{Approximating bribery in scoring rules},
\newblock in: \bibinfo{editor}{S.~A. McIlraith}, \bibinfo{editor}{K.~Q. Weinberger} (Eds.), \bibinfo{booktitle}{Proceedings of the Thirty-Second {AAAI} Conference on Artificial Intelligence, (AAAI-18), the 30th innovative Applications of Artificial Intelligence (IAAI-18), and the 8th {AAAI} Symposium on Educational Advances in Artificial Intelligence (EAAI-18), New Orleans, Louisiana, USA, February 2-7, 2018}, \bibinfo{publisher}{{AAAI} Press}, \bibinfo{year}{2018}, pp. \bibinfo{pages}{1121--1129}. \URLprefix \url{https://doi.org/10.1609/aaai.v32i1.11476}. \DOIprefix\doi{10.1609/AAAI.V32I1.11476}.
\bibitem[{Liu et~al.(2009)Liu, Feng, Zhu, and Luan}]{liu2009parameterized}
\bibinfo{author}{H.~Liu}, \bibinfo{author}{H.~Feng}, \bibinfo{author}{D.~Zhu}, \bibinfo{author}{J.~Luan},
\newblock \bibinfo{title}{Parameterized computational complexity of control problems in voting systems},
\newblock \bibinfo{journal}{Theoretical Computer Science} \bibinfo{volume}{410} (\bibinfo{year}{2009}) \bibinfo{pages}{2746--2753}.
\bibitem[{Tao et~al.(2022)Tao, Chen, Xu, Shi, Sunny, and Uz~Zaman}]{tao2022hard}
\bibinfo{author}{L.~Tao}, \bibinfo{author}{L.~Chen}, \bibinfo{author}{L.~Xu}, \bibinfo{author}{W.~Shi}, \bibinfo{author}{A.~Sunny}, \bibinfo{author}{M.~M. Uz~Zaman},
\newblock \bibinfo{title}{How hard is bribery in elections with randomly selected voters},
\newblock in: \bibinfo{booktitle}{Proceedings of the 21st International Conference on Autonomous Agents and Multiagent Systems}, \bibinfo{year}{2022}.
\bibitem[{Yang(2019)}]{Yang2019Complexity}
\bibinfo{author}{Y.~Yang},
\newblock \bibinfo{title}{Complexity of manipulating and controlling approval-based multiwinner voting},
\newblock in: \bibinfo{booktitle}{Proceedings of the Twenty-Eighth International Joint Conference on Artificial Intelligence, {IJCAI-19}}, \bibinfo{publisher}{International Joint Conferences on Artificial Intelligence Organization}, \bibinfo{year}{2019}, pp. \bibinfo{pages}{637--643}. \URLprefix \url{https://doi.org/10.24963/ijcai.2019/90}. \DOIprefix\doi{10.24963/ijcai.2019/90}.
\bibitem[{Zhou and Guo(2020)}]{zhou2020parameterized}
\bibinfo{author}{A.~Zhou}, \bibinfo{author}{J.~Guo},
\newblock \bibinfo{title}{Parameterized complexity of shift bribery in iterative elections},
\newblock in: \bibinfo{booktitle}{Proceedings of the 19th International Conference on Autonomous Agents and Multiagent Systems}, \bibinfo{year}{2020}, pp. \bibinfo{pages}{1665--1673}.
\bibitem[{Bartholdi~III et~al.(1992)Bartholdi~III, Tovey, and Trick}]{bartholdi1992hard}
\bibinfo{author}{J.~J. Bartholdi~III}, \bibinfo{author}{C.~A. Tovey}, \bibinfo{author}{M.~A. Trick},
\newblock \bibinfo{title}{How hard is it to control an election?},
\newblock \bibinfo{journal}{Mathematical and Computer Modelling} \bibinfo{volume}{16} (\bibinfo{year}{1992}) \bibinfo{pages}{27--40}.
\bibitem[{Brelsford et~al.(2008)Brelsford, Faliszewski, Hemaspaandra, Schnoor, and Schnoor}]{brelsford2008approximability}
\bibinfo{author}{E.~Brelsford}, \bibinfo{author}{P.~Faliszewski}, \bibinfo{author}{E.~Hemaspaandra}, \bibinfo{author}{H.~Schnoor}, \bibinfo{author}{I.~Schnoor},
\newblock \bibinfo{title}{Approximability of manipulating elections},
\newblock in: \bibinfo{editor}{D.~Fox}, \bibinfo{editor}{C.~P. Gomes} (Eds.), \bibinfo{booktitle}{Proceedings of the Twenty-Third {AAAI} Conference on Artificial Intelligence, {AAAI} 2008, Chicago, Illinois, USA, July 13-17, 2008}, \bibinfo{publisher}{{AAAI} Press}, \bibinfo{year}{2008}, pp. \bibinfo{pages}{44--49}. \URLprefix \url{http://www.aaai.org/Library/AAAI/2008/aaai08-007.php}.
\bibitem[{Faliszewski(2008)}]{faliszewski2008Nonuniform}
\bibinfo{author}{P.~Faliszewski},
\newblock \bibinfo{title}{Nonuniform bribery},
\newblock in: \bibinfo{editor}{L.~Padgham}, \bibinfo{editor}{D.~C. Parkes}, \bibinfo{editor}{J.~P. M{\"{u}}ller}, \bibinfo{editor}{S.~Parsons} (Eds.), \bibinfo{booktitle}{7th International Joint Conference on Autonomous Agents and Multiagent Systems {(AAMAS} 2008), Estoril, Portugal, May 12-16, 2008, Volume 3}, \bibinfo{publisher}{{IFAAMAS}}, \bibinfo{year}{2008}, pp. \bibinfo{pages}{1569--1572}. \URLprefix \url{https://dl.acm.org/citation.cfm?id=1402927}.
\bibitem[{Faliszewski et~al.(2015)Faliszewski, Reisch, Rothe, and Schend}]{faliszewski2015complexity}
\bibinfo{author}{P.~Faliszewski}, \bibinfo{author}{Y.~Reisch}, \bibinfo{author}{J.~Rothe}, \bibinfo{author}{L.~Schend},
\newblock \bibinfo{title}{Complexity of manipulation, bribery, and campaign management in bucklin and fallback voting},
\newblock \bibinfo{journal}{Autonomous Agents and Multi-Agent Systems} \bibinfo{volume}{29} (\bibinfo{year}{2015}) \bibinfo{pages}{1091--1124}.
\bibitem[{Maushagen et~al.(2022)Maushagen, Neveling, Rothe, and Selker}]{maushagen2022complexity}
\bibinfo{author}{C.~Maushagen}, \bibinfo{author}{M.~Neveling}, \bibinfo{author}{J.~Rothe}, \bibinfo{author}{A.-K. Selker},
\newblock \bibinfo{title}{Complexity of shift bribery for iterative voting rules},
\newblock \bibinfo{journal}{Annals of Mathematics and Artificial Intelligence} \bibinfo{volume}{90} (\bibinfo{year}{2022}) \bibinfo{pages}{1017--1054}.
\bibitem[{Parkes and Xia(2012)}]{parkes2012complexity}
\bibinfo{author}{D.~Parkes}, \bibinfo{author}{L.~Xia},
\newblock \bibinfo{title}{A complexity-of-strategic-behavior comparison between schulze's rule and ranked pairs},
\newblock in: \bibinfo{booktitle}{Proceedings of the AAAI Conference on Artificial Intelligence}, volume~\bibinfo{volume}{26}, \bibinfo{year}{2012}, pp. \bibinfo{pages}{1429--1435}.
\bibitem[{Menton and Singh(2013)}]{menton2013control}
\bibinfo{author}{C.~G. Menton}, \bibinfo{author}{P.~Singh},
\newblock \bibinfo{title}{Control complexity of schulze voting},
\newblock in: \bibinfo{booktitle}{Twenty-Third International Joint Conference on Artificial Intelligence}, \bibinfo{year}{2013}.
\bibitem[{Procaccia et~al.(2007)Procaccia, Rosenschein, and Zohar}]{procaccia2007multi}
\bibinfo{author}{A.~D. Procaccia}, \bibinfo{author}{J.~S. Rosenschein}, \bibinfo{author}{A.~Zohar},
\newblock \bibinfo{title}{Multi-winner elections: Complexity of manipulation, control and winner-determination.},
\newblock in: \bibinfo{booktitle}{IJCAI}, volume~\bibinfo{volume}{7}, \bibinfo{year}{2007}, pp. \bibinfo{pages}{1476--1481}.
\bibitem[{Put and Faliszewski(2016)}]{put2016complexity}
\bibinfo{author}{T.~Put}, \bibinfo{author}{P.~Faliszewski},
\newblock \bibinfo{title}{The complexity of voter control and shift bribery under parliament choosing rules},
\newblock \bibinfo{journal}{Trans. Comput. Collect. Intell.} \bibinfo{volume}{23} (\bibinfo{year}{2016}) \bibinfo{pages}{29--50}. \URLprefix \url{https://doi.org/10.1007/978-3-662-52886-0\_3}. \DOIprefix\doi{10.1007/978-3-662-52886-0\_3}.
\bibitem[{Sina et~al.(2015)Sina, Hazon, Hassidim, and Kraus}]{sina2015adapting}
\bibinfo{author}{S.~Sina}, \bibinfo{author}{N.~Hazon}, \bibinfo{author}{A.~Hassidim}, \bibinfo{author}{S.~Kraus},
\newblock \bibinfo{title}{Adapting the social network to affect elections},
\newblock in: \bibinfo{booktitle}{Proceedings of the 2015 international conference on autonomous agents and multiagent systems}, \bibinfo{organization}{Citeseer}, \bibinfo{year}{2015}, pp. \bibinfo{pages}{705--713}.
\bibitem[{Maushagen and Rothe(2020)}]{maushagen2020last}
\bibinfo{author}{C.~Maushagen}, \bibinfo{author}{J.~Rothe},
\newblock \bibinfo{title}{The last voting rule is home: Complexity of control by partition of candidates or voters in maximin elections},
\newblock in: \bibinfo{booktitle}{ECAI 2020}, \bibinfo{publisher}{IOS Press}, \bibinfo{year}{2020}, pp. \bibinfo{pages}{163--170}.
\bibitem[{Meir et~al.(2008)Meir, Procaccia, Rosenschein, and Zohar}]{meir2008complexity}
\bibinfo{author}{R.~Meir}, \bibinfo{author}{A.~D. Procaccia}, \bibinfo{author}{J.~S. Rosenschein}, \bibinfo{author}{A.~Zohar},
\newblock \bibinfo{title}{Complexity of strategic behavior in multi-winner elections},
\newblock \bibinfo{journal}{Journal of Artificial Intelligence Research} \bibinfo{volume}{33} (\bibinfo{year}{2008}) \bibinfo{pages}{149--178}.
\bibitem[{Aziz et~al.(2015)Aziz, Gaspers, Gudmundsson, Mackenzie, Mattei, and Walsh}]{aziz2015computational}
\bibinfo{author}{H.~Aziz}, \bibinfo{author}{S.~Gaspers}, \bibinfo{author}{J.~Gudmundsson}, \bibinfo{author}{S.~Mackenzie}, \bibinfo{author}{N.~Mattei}, \bibinfo{author}{T.~Walsh},
\newblock \bibinfo{title}{Computational aspects of multi-winner approval voting.},
\newblock in: \bibinfo{booktitle}{AAMAS}, volume~\bibinfo{volume}{15}, \bibinfo{year}{2015}, pp. \bibinfo{pages}{107--115}.
\bibitem[{Bredereck et~al.(2016)Bredereck, Faliszewski, Niedermeier, and Talmon}]{bredereck2016complexity}
\bibinfo{author}{R.~Bredereck}, \bibinfo{author}{P.~Faliszewski}, \bibinfo{author}{R.~Niedermeier}, \bibinfo{author}{N.~Talmon},
\newblock \bibinfo{title}{Complexity of shift bribery in committee elections},
\newblock in: \bibinfo{booktitle}{Proceedings of the AAAI Conference on Artificial Intelligence}, volume~\bibinfo{volume}{30}, \bibinfo{year}{2016}.
\bibitem[{Bredereck et~al.(2021)Bredereck, Kaczmarczyk, and Niedermeier}]{bredereck2021coalitional}
\bibinfo{author}{R.~Bredereck}, \bibinfo{author}{A.~Kaczmarczyk}, \bibinfo{author}{R.~Niedermeier},
\newblock \bibinfo{title}{On coalitional manipulation for multiwinner elections: Shortlisting},
\newblock \bibinfo{journal}{Autonomous Agents and Multi-Agent Systems} \bibinfo{volume}{35} (\bibinfo{year}{2021}) \bibinfo{pages}{38}.
\bibitem[{Obraztsova et~al.(2013)Obraztsova, Zick, and Elkind}]{obraztsova2013manipulation}
\bibinfo{author}{S.~Obraztsova}, \bibinfo{author}{Y.~Zick}, \bibinfo{author}{E.~Elkind},
\newblock \bibinfo{title}{On manipulation in multi-winner elections based on scoring rules},
\newblock in: \bibinfo{booktitle}{Proceedings of the 2013 international conference on Autonomous agents and multi-agent systems}, \bibinfo{year}{2013}, pp. \bibinfo{pages}{359--366}.
\bibitem[{Yang(2025)}]{Yang2025Destructive}
\bibinfo{author}{Y.~Yang},
\newblock \bibinfo{title}{On the complexity of destructive bribery in approval-based multiwinner voting},
\newblock \bibinfo{journal}{ACM Trans. Comput. Theory}  (\bibinfo{year}{2025}). \URLprefix \url{https://doi.org/10.1145/3749377}. \DOIprefix\doi{10.1145/3749377}, \bibinfo{note}{just Accepted}.
\bibitem[{Taylor and Zwicker(2021)}]{taylor2021simple}
\bibinfo{author}{A.~D. Taylor}, \bibinfo{author}{W.~S. Zwicker},
\newblock \bibinfo{title}{Simple games: Desirability relations, trading, pseudoweightings}  (\bibinfo{year}{2021}).
\bibitem[{Aziz et~al.(2011)Aziz, Bachrach, Elkind, and Paterson}]{Aziz2011False}
\bibinfo{author}{H.~Aziz}, \bibinfo{author}{Y.~Bachrach}, \bibinfo{author}{E.~Elkind}, \bibinfo{author}{M.~Paterson},
\newblock \bibinfo{title}{False-name manipulations in weighted voting games},
\newblock \bibinfo{journal}{J. Artif. Int. Res.} \bibinfo{volume}{40} (\bibinfo{year}{2011}) \bibinfo{pages}{57–93}.
\bibitem[{Rey and Rothe(2014)}]{rey2014false}
\bibinfo{author}{A.~Rey}, \bibinfo{author}{J.~Rothe},
\newblock \bibinfo{title}{False-name manipulation in weighted voting games is hard for probabilistic polynomial time},
\newblock \bibinfo{journal}{Journal of Artificial Intelligence Research} \bibinfo{volume}{50} (\bibinfo{year}{2014}) \bibinfo{pages}{573--601}.
\bibitem[{Rey and Rothe(2018)}]{rey2018structural}
\bibinfo{author}{A.~Rey}, \bibinfo{author}{J.~Rothe},
\newblock \bibinfo{title}{Structural control in weighted voting games},
\newblock \bibinfo{journal}{The BE Journal of Theoretical Economics} \bibinfo{volume}{18} (\bibinfo{year}{2018}) \bibinfo{pages}{20160169}.
\bibitem[{Zick et~al.(2011)Zick, Skopalik, and Elkind}]{zick2011shapley}
\bibinfo{author}{Y.~Zick}, \bibinfo{author}{A.~Skopalik}, \bibinfo{author}{E.~Elkind},
\newblock \bibinfo{title}{The shapley value as a function of the quota in weighted voting games},
\newblock in: \bibinfo{booktitle}{22nd International Joint Conference on Artificial Intelligence, IJCAI 2011}, \bibinfo{year}{2011}, pp. \bibinfo{pages}{490--495}.
\bibitem[{Zuckerman et~al.(2012)Zuckerman, Faliszewski, Bachrach, and Elkind}]{zuckerman2012manipulating}
\bibinfo{author}{M.~Zuckerman}, \bibinfo{author}{P.~Faliszewski}, \bibinfo{author}{Y.~Bachrach}, \bibinfo{author}{E.~Elkind},
\newblock \bibinfo{title}{Manipulating the quota in weighted voting games},
\newblock \bibinfo{journal}{Artificial Intelligence} \bibinfo{volume}{180} (\bibinfo{year}{2012}) \bibinfo{pages}{1--19}.
\bibitem[{Slinko and White(2010)}]{slinko2010proportional}
\bibinfo{author}{A.~Slinko}, \bibinfo{author}{S.~White},
\newblock \bibinfo{title}{Proportional representation and strategic voters},
\newblock \bibinfo{journal}{Journal of Theoretical Politics} \bibinfo{volume}{22} (\bibinfo{year}{2010}) \bibinfo{pages}{301--332}.
\bibitem[{Bowler et~al.(2010)Bowler, Karp, and Donovan}]{bowler2010strategic}
\bibinfo{author}{S.~Bowler}, \bibinfo{author}{J.~A. Karp}, \bibinfo{author}{T.~Donovan},
\newblock \bibinfo{title}{Strategic coalition voting: evidence from new zealand},
\newblock \bibinfo{journal}{Electoral Studies} \bibinfo{volume}{29} (\bibinfo{year}{2010}) \bibinfo{pages}{350--357}.
\bibitem[{Cox(2018)}]{cox2018portfolio}
\bibinfo{author}{G.~W. Cox},
\newblock \bibinfo{title}{Portfolio-maximizing strategic voting in parliamentary elections},
\newblock \bibinfo{journal}{The Oxford handbook of electoral systems}  (\bibinfo{year}{2018}) \bibinfo{pages}{265}.
\bibitem[{Gschwend et~al.(2016)Gschwend, Stoetzer, and Zittlau}]{gschwend2016drives}
\bibinfo{author}{T.~Gschwend}, \bibinfo{author}{L.~Stoetzer}, \bibinfo{author}{S.~Zittlau},
\newblock \bibinfo{title}{What drives rental votes? how coalitions signals facilitate strategic coalition voting},
\newblock \bibinfo{journal}{Electoral Studies} \bibinfo{volume}{44} (\bibinfo{year}{2016}) \bibinfo{pages}{293--306}.
\bibitem[{McCuen and Morton(2010)}]{MCCUEN2010316}
\bibinfo{author}{B.~McCuen}, \bibinfo{author}{R.~B. Morton},
\newblock \bibinfo{title}{Tactical coalition voting and information in the laboratory},
\newblock \bibinfo{journal}{Electoral Studies} \bibinfo{volume}{29} (\bibinfo{year}{2010}) \bibinfo{pages}{316--328}. \URLprefix \url{https://www.sciencedirect.com/science/article/pii/S0261379410000211}. \DOIprefix\doi{https://doi.org/10.1016/j.electstud.2010.03.009}, \bibinfo{note}{special Symposium: Voters and Coalition Governments}.
\bibitem[{Meffert and Gschwend(2010)}]{meffert2010strategic}
\bibinfo{author}{M.~F. Meffert}, \bibinfo{author}{T.~Gschwend},
\newblock \bibinfo{title}{Strategic coalition voting: Evidence from austria},
\newblock \bibinfo{journal}{Electoral Studies} \bibinfo{volume}{29} (\bibinfo{year}{2010}) \bibinfo{pages}{339--349}.
\bibitem[{Boratyn et~al.(2024)Boratyn, Słomczyński, and Stolicki}]{boratyn2024seatallocationseatbias}
\bibinfo{author}{D.~Boratyn}, \bibinfo{author}{W.~Słomczyński}, \bibinfo{author}{D.~Stolicki}, \bibinfo{title}{Seat allocation and seat bias under the jefferson--d'hondt method}, \bibinfo{year}{2024}. \URLprefix \url{https://arxiv.org/abs/1805.08291}. \href{http://arxiv.org/abs/1805.08291}{{\tt arXiv:1805.08291}}.
\bibitem[{Flis et~al.(2020)Flis, S{\l}omczy{\'n}ski, and Stolicki}]{FlisEtAl2020}
\bibinfo{author}{J.~Flis}, \bibinfo{author}{W.~S{\l}omczy{\'n}ski}, \bibinfo{author}{D.~Stolicki},
\newblock \bibinfo{title}{Pot and ladle: a formula for estimating the distribution of seats under the jefferson–d'hondt method},
\newblock \bibinfo{journal}{Public Choice} \bibinfo{volume}{182} (\bibinfo{year}{2020}) \bibinfo{pages}{201--227}. \DOIprefix\doi{10.1007/s11127-019-00680-w}.
\bibitem[{Ahuja et~al.(1993)Ahuja, Magnanti, and Orlin}]{ahuja1993network}
\bibinfo{author}{R.~K. Ahuja}, \bibinfo{author}{T.~L. Magnanti}, \bibinfo{author}{J.~B. Orlin},
\newblock \bibinfo{title}{Network flows: Theory, applications and algorithms},
\newblock \bibinfo{journal}{Englewood Cliffs, New Jersey, USA Arrow, KJ: Prentice-Hall}  (\bibinfo{year}{1993}).
\bibitem[{Downey et~al.(1999)Downey, Fellows, Downey, and Fellows}]{downey1999structure}
\bibinfo{author}{R.~Downey}, \bibinfo{author}{M.~Fellows}, \bibinfo{author}{R.~Downey}, \bibinfo{author}{M.~Fellows},
\newblock \bibinfo{title}{The structure of languages under parameterized reducibilities},
\newblock \bibinfo{journal}{Parameterized Complexity}  (\bibinfo{year}{1999}) \bibinfo{pages}{389--437}.
\bibitem[{Downey and Fellows(1995)}]{Fixed1995Downey}
\bibinfo{author}{R.~G. Downey}, \bibinfo{author}{M.~R. Fellows},
\newblock \bibinfo{title}{Fixed-parameter tractability and completeness i: Basic results},
\newblock \bibinfo{journal}{SIAM Journal on Computing} \bibinfo{volume}{24} (\bibinfo{year}{1995}) \bibinfo{pages}{873--921}. \DOIprefix\doi{10.1137/S0097539792228228}.

\end{thebibliography}
